\documentclass{article}
\usepackage{braket}
\usepackage{comment}
\usepackage{amssymb}
\usepackage{graphicx}
\usepackage{amsmath}
\usepackage{array}
\usepackage{longtable}
\usepackage{booktabs}
\usepackage{tabularx}
\usepackage{adjustbox}
\usepackage{float}       
\usepackage{placeins}    
\usepackage{xcolor}
\usepackage{subcaption}
\usepackage[colorlinks=true, allcolors=blue]{hyperref}
\usepackage[top=2cm,bottom=2cm,left=2cm,right=2cm]{geometry}
\usepackage{listings}
\usepackage{enumitem}
\usepackage[normalem]{ulem}
\usepackage{bm}
\usepackage{tikz} 
\usetikzlibrary{matrix, arrows.meta, positioning, calc, decorations.markings}
\usepackage{tikz-cd}
\usepackage{mathtools}
\usepackage{ulem}
\usepackage{dsfont}
\usepackage{cancel}
\tikzcdset{
  cells={nodes={draw,rounded corners,fill=gray!10,inner xsep=6pt,inner ysep=4pt}},
  arrows={>=latex}
}

\title{Implementation and Analysis of Quantum Majority Rules under Noisy Conditions}
\author{
  Gal Amit$^{1}$,
  Yuval Idan$^{1}$,
  Michael Suleymanov$^{1}$,
  Luis Razo$^{2}$, and
  Eliahu Cohen$^{1,\ast}$\\[0.5em]
  {\small $^{1}$Faculty of Engineering and the Institute of Nanotechnology and Advanced Materials, Bar-Ilan University, Ramat Gan 5290002, Israel}\\
  {\small $^{2}$European Institute of Science in Management, Barcelona 08036, Spain}\\[0.5em]
  {\small $^{\ast}$Corresponding author: {eliahu.cohen@biu.ac.il}}
}

\date{}

\begin{document}

\maketitle

\begin{abstract}
Quantum voting, inspired by quantum game theory, provides a framework in which the quantum majority rule (QMR) constitution of Bao and Yunger Halpern [Phys. Rev. A 95, 062306 (2017)] violates the quantum analogue of Arrow’s impossibility theorem. We evaluate this QMR constitution analytically on classical profile data and implement its final measurement stage as a quantum circuit, running on both noiseless simulators and noisy IBM quantum hardware to map how realistic noise deforms the resulting societal ranking distribution. Moderate–high single-qubit noise does not change the qualitative behavior of QMR, whereas strong noise shifts the distribution toward other dominant winners than the classical one. We quantify this
behavior using winner–agreement rates, Condorcet-winner flip rates, and
Jensen–Shannon divergence between societal ranking distributions.

In a second, exploratory component, we demonstrate an explicitly entanglement-based variant of the QMR constitution that serves as a testbed for multi-voter quantum correlations under noise, which we refer to as the QMR2-inspired variant. There, GHZ-type and separable superpositions over opposite rankings have the same expectation values but respond very differently to noise.
Taken together, these two components connect the abstract QMR constitution to concrete implementations on noisy intermediate-scale quantum (NISQ) devices and highlight design considerations for future quantum voting protocols.

\end{abstract}


\section{Introduction}

Voting protocols are central to collective decision-making, with applications ranging from political elections to distributed multi-agent systems and social networks. At their mathematical core, voting systems can be studied within the broad discipline of game theory, which seeks to model and analyze the strategic behavior of rational agents in situations of conflict or cooperation. Classical game theory, formalized by John von Neumann and Oskar Morgenstern and later extended by John Nash, has produced a deep understanding of how individual preferences, strategies, and payoffs interact to yield collective outcomes~\cite{vonNeumann1944,Nash1951}.

This understanding has found profound applications not only in economics and political science but also in distributed computation, evolutionary biology, and computer security.

\subsection{Classical and Quantum Game-Theoretic Foundations of Social Choice and Voting}

Classical game theory provides a rigorous language for analyzing strategic interactions among rational agents. In the context of voting, these agents--voters possess individual preferences over a set of alternatives and are often modeled as self-interested, strategic players. Classical voting systems, such as plurality voting, Borda count, and Condorcet methods, are designed to aggregate these individual preferences into a collective decision. A concise summary of the notation and basic definitions used for classical ranked voting is collected in Appendix~\ref{app_classical_choice_theory__notations}. Yet, despite the sophistication of classical models, social choice theory quickly encounters intrinsic barriers.

The most widely used single-winner voting scheme is the plurality scheme, where voters select one candidate and the candidate with the most votes wins. Plurality is vulnerable to issues such as vote splitting, the spoiler effect, lack of majority support, and strategic voting. 
Ranked schemes—where voters submit a preference order—can mitigate some of these effects by capturing more information than a single choice. For example, instant-runoff voting (IRV) can reduce spoiler dynamics and wasted votes relative to plurality, though it does not eliminate strategic behavior or paradoxes in general, and it may disagree with Condorcet-consistent rules on some profiles.

This result is not merely a technicality but rather a deep statement about the inherent trade-offs in designing collective decision-making mechanisms. As a consequence, any classical voting protocol is unavoidably subject to some form of inconsistency, vulnerability to strategic manipulation, or outright unfairness.

Research has thus sought to chart the limits and possibilities within these constraints. The Gibbard-Satterthwaite theorem further deepens the pessimism by showing that, under broad conditions, every non-dictatorial and onto voting rule is susceptible to tactical manipulation by voters~\cite{Gibbard1973, Satterthwaite1975}. These impossibility results have catalyzed decades of research into voting systems that seek to optimize specific trade-offs among fairness, robustness, and resistance to manipulation.

Nevertheless, the ultimate limitations of classical voting remain starkly evident: the strategic power of rational agents and the combinatorial complexity of preference aggregation conspire to make ``perfect'' voting unattainable within the classical framework. For completeness, the formal statement of Arrow's axioms and a compact presentation of the impossibility theorem are collected in Appendix~\ref{app_Arrow's_properties}.

Game-theoretic models have also proved foundational for the analysis of so-called nonlocal games, which provide a vivid operational setting for exploring the limits of classical and quantum strategies~\cite{Cleve2004, Brassard2005}. In such games, two or more players cooperate to achieve a common goal but are forbidden from communicating during play. Classical strategies are constrained by local realism---players can agree on a common strategy in advance but cannot use any form of instantaneous or nonlocal coordination. These models have deep analogies with voting protocols, where the aggregation of independent inputs into a collective outcome is subject to similar informational constraints.

The emergence of quantum information science has radically transformed our understanding of both computation and strategic interaction. Quantum mechanics introduces phenomena---most notably superposition and entanglement---that allow agents to transcend classical limitations. Quantum game theory extends the classical paradigm by allowing players to encode strategies in quantum states, employ entangled resources, and perform measurements that admit correlations impossible within any local realistic theory~\cite{Cleve2004, Brassard2005, Meyer1999, Eisert1999}.

The power of quantum strategies is particularly manifest in nonlocal games, where the use of entanglement allows players to achieve winning probabilities strictly exceeding those attainable by any classical strategy. The seminal Clauser--Horne--Shimony--Holt (CHSH) game~\cite{CHSH1969} illustrates this vividly: two cooperating players, Alice and Bob, receive correlated questions and must output binary answers without communicating. If their shared resource is classical, the optimal success probability is $0.75$, whereas an entangled strategy attains $\cos^2(\pi/8)\approx 0.854$~\cite{Cleve2004, Brassard2005}. Such violations of Bell inequalities directly demonstrate quantum nonlocality and reveal how quantum resources enlarge the space of achievable correlations.

Beyond CHSH, nonlocal games now provide a broad framework for probing the limits of classical and quantum correlations. Quantum games are distinguished by their allowance for quantum strategies: agents can share entangled states, perform joint measurements, and exploit quantum correlations that cannot be simulated by any local hidden-variable model. These features allow for striking phenomena such as pseudo-telepathy~\cite{Brassard2005}, in which players using quantum resources can win certain games with certainty, whereas classical players always lose with nonzero probability. A canonical example is the magic square game, where quantum players can always satisfy the game’s consistency constraints, while classical players cannot~\cite{Cabello2001, Aravind2004}. However, not all nonlocal games admit perfect quantum strategies; for binary-output games such as CHSH and odd-cycle games, the quantum advantage is quantitatively bounded~\cite{Brassard2005}.

Experimentally, realizing quantum advantages in nonlocal games is challenging, due to requirements on detection efficiencies, spacelike separation, and the closing of loopholes. Early demonstrations based on entangled photons were limited by detector inefficiencies and the notorious detection loophole~\cite{Xu2022, Christensen2015}. More recent experiments with trapped ions and state-of-the-art photonic platforms have substantially tightened these constraints, achieving steadily improving detection efficiencies and spacelike separation protocols. These advances provide increasingly rigorous tests of quantum nonlocality and pave the way toward scalable, multi-party implementations relevant to collective decision-making.

In the context of social choice, quantum voting protocols generalize classical preference aggregation by encoding voters’ preferences into quantum states---often as superpositions or mixtures of classical rankings. The aggregation process can then leverage quantum coherence and entanglement to realize new forms of collective decision-making. Unlike probabilistic mixtures of classical rankings, quantum superpositions enable interference effects that have no classical analog, which in principle allows one to circumvent some impossibility results such as Arrow’s theorem in suitably generalized, quantum formulations. For reference, the basic quantum-mechanical notation used in our quantum voting framework is summarized in Table~\ref{tab:qm_notation} in the appendices.

A paradigmatic instance of this idea is the protocol of Bao and Yunger  Halpern (BYH)~\cite{BaoHalpern2017}. Their quantum majority rule (QMR) constitution encodes voter preferences as superpositions over classical rankings and aggregates them via quantum operations followed by classical post-processing. In this framework, QMR can violate a quantum analogue of Arrow’s impossibility theorem in a precisely defined setting, demonstrating how quantum coherence and superposition can alter fundamental limitations of classical social choice.

These developments suggest that quantum resources can modify strategic interaction in ways unattainable classically and motivate the construction of concrete quantum voting schemes. In particular, implementing the BYH QMR constitution on realistic, noisy intermediate-scale quantum (NISQ) hardware offers a natural test bed for studying how noise, decoherence, and device imperfections deform the intended societal preference distribution. The present work takes up this challenge by providing a full algorithmic specification of QMR and investigating its stability under experimentally relevant noise models and hardware conditions.

\subsection{Violation of Arrow's theorem, modified in the quantum spirit}\label{subsec_BYH}

A key breakthrough, due to BYH~\cite{BaoHalpern2017},
can be organized into three conceptual steps:
\begin{itemize}
    \item Association of society with a joint quantum state $\sigma_\text{soc}$ that may include non-classical correlations between voters' preferences.
    \item Formulation of the quantum version of Arrow’s properties (see Appendix~\ref{app_BYH_quantum_Arrow's_properties} for more details).
    \item Construction of a quantum voting protocol that violates the quantum modification of Arrow’s impossibility theorem (see Appendix~\ref{app_BYH_quantum_QMR} for more details).
\end{itemize}
In the following, we briefly summarize the idea proposed in~\cite{BaoHalpern2017} (for more details see Table~\ref{tab_QMR1_BYH}).
For $n$ voters, $V=\{1,\ldots,n\}$, and $m$ alternatives $A=\{a_1,\ldots,a_m\}$, the society's joint quantum state is assumed to belong to the density operators space, $\sigma_\text{soc}\in \mathcal{D}(\mathcal{H}_A^{\otimes n})$, where $\mathcal{H}_A$ is the Hilbert space associated with a single voter's preference over the alternatives set $A$.  Quantum Social Welfare function (QSWF), constitution, is a completely positive trace-preserving (CPTP) map, $\mathcal{E}(\sigma_\text{soc})=\rho_\text{soc}\in\mathcal{D}(\mathcal{H})$.
The main constitution described in BYH's paper is called the QMR. 
Each voter's preference may be obtained by tracing out all the rest, $\rho_i=\text{Tr}_{\ne i}(\sigma_\text{soc})$, and the set of all voters' preferences is called the quantum profile
$\mathcal{P}_\text{soc}=\text{QP}(\sigma_\text{soc})=\{\rho_1,\ldots,\rho_N\}$,
where QP denotes the quantum-profile map that associates a joint state with the corresponding list of single-voter reduced states.

The next step in $\mathcal{E}_\text{QMR}$ constitution is the decoherence of the
voter's preferences\footnote{By voters' preferences, here we mean strict linear orders of the form $L_i=(a_1\succ a_2\cdots\succ a_m)\in\mathcal{L}(A)$, where $i\in V$, and $a_i\in A$. For each pair, $a,b\in A$, and each voter, $i\in V$, we have either $a\succ_i b$ or $b\succ_i a$.} in the preference basis $\{\ket{L}\}$, namely,
$\rho_i^d=\sum_{L_i}\ket{L_i}\bra{L_i}\rho_i\ket{L_i}
\bra{L_i}\equiv\sum_{L_i}p_i^{L_i}\chi_i^{L_i}$, where
$L_i\in\mathcal{L}(A)$ are strict linear orders,
$p_i^{L_i}=\bra{L_i}\rho_i\ket{L_i}$, and
$\chi_i^{L_i}=\ket{L_i}\bra{L_i}$. This gives a dephased quantum
profile $\mathcal{P}^d_\text{soc}=\text{DP}(\mathcal{P}_\text{soc})=
\{\rho_1^d,\ldots,\rho_n^d\}$, where DP denotes the dephasing map that removes
off-diagonal coherences in the preference basis.

The resulting society's dephased product state of the form
\begin{equation}
  \sigma^{\text{DP}}_\text{soc}=\text{PS}(\mathcal{P}^d_\text{soc})
=\bigotimes_{i=1}^n\rho_i^d=
\sum_{L_1,\ldots,L_n}p_1^{L_1}\cdots p_n^{L_n}
\bigotimes_{i=1}^n\chi_i^{L_i}
\equiv 
\sum_{\vec{L}} p(\vec{L}) \vec{\chi}(\vec{L}),
\end{equation}
may be constructed, where 
\begin{equation}
    \vec{L}=(L_1,\ldots,L_n),\hspace{3mm}
    p(\vec{L})=p_1^{L_1}\cdots p_n^{L_n},\hspace{3mm} 
    \vec{\chi}(\vec{L})=\bigotimes_{i=1}^n\chi_i^{L_i}
\end{equation}
here PS denotes
the product-state map that forms the tensor product of the dephased single-voter
states.

For each $\vec{\chi}(\vec{L})$, $\mathcal{E}_\text{QMR}$ constructs a directed graph for which the list of strongly connected components (SCCs) is calculated using Tarjan's algorithm~\cite{Tarjan1972}. 
In this step, $\mathcal{E}_\text{QMR}$ outputs the uniform mixed state over all linear extensions ($\mathcal{L}_\text{SCC}$) of the partial order induced by the SCCs, $\chi^{(1)}(\vec{L})=\text{DS}\left(
\vec{\chi}(\vec{L})
\right)=\frac{1}{|\mathcal{L}_\text{SCC}|}\sum_{L\in \mathcal{L}_\text{SCC}}\ket{L}\bra{L}$, i.e., the maximally mixed state on the subspace spanned by those preference-basis elements.
The next step, called \textit{give the minority a shot} (GMS), increases the weights of preferences not represented in $\chi^{(1)}(\vec{L})$ by a factor $\delta$, giving $\chi^{(2)}(\vec{L})=\text{GMS}_\delta\left(\chi^{(1)}(\vec{L})\right)$. The GMS step must be followed by a compensation step of enforcing unanimity (EU), to satisfy the mandatory properties, $\chi_\text{soc}(\vec{L})=\text{EU}\left(\chi^{(2)}(\vec{L})\right)$. 
Combining all the steps, the QMR1 SWF is of the form 
$\mathcal{E}_\text{QMR}=\text{EU}\circ \text{GMS} \circ\text{DS}\circ\text{PS}\circ\text{DP}\circ\text{QP}$.
The resulting state, $\rho_\text{soc}=\mathcal{E}_\text{QMR}(\sigma_\text{soc})=
\sum_{\vec{L}} p(\vec{L})\chi_\text{soc}(\vec{L})$, may finally be measured in the preferences basis, $\{\ket{L}\}$, to provide the single outcome of the quantum election.

\begin{figure}[H]
    \centering
    \includegraphics[width=1\linewidth]{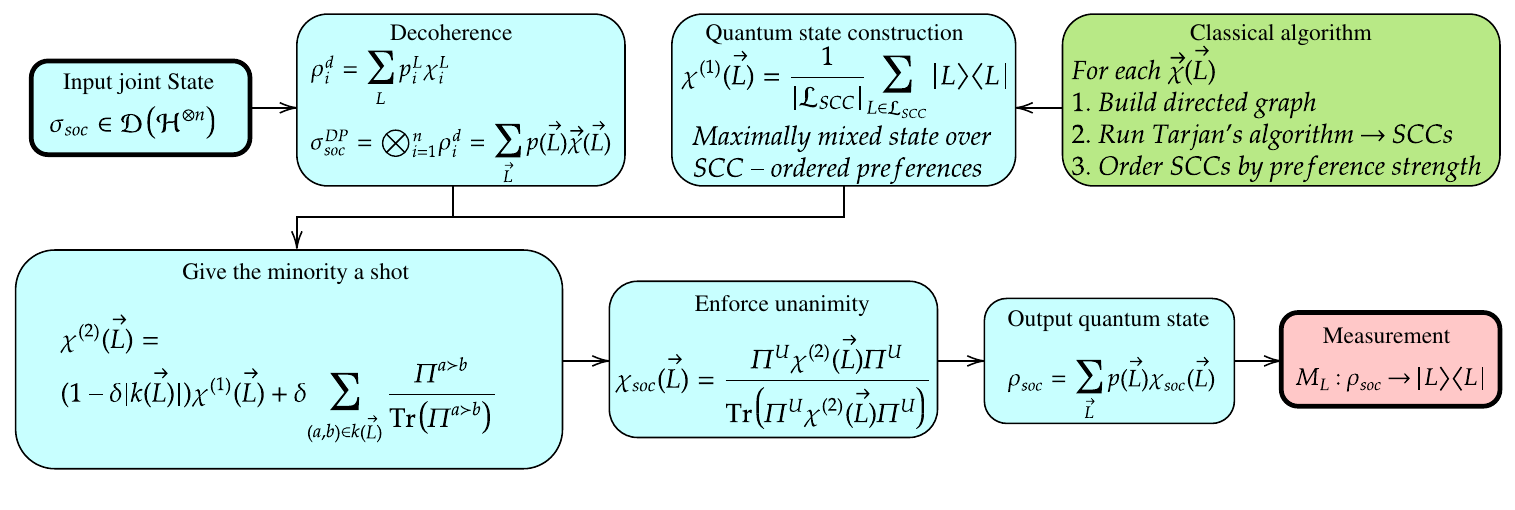}
    \caption{{\bf Quantum Majority rule.} Streamlined version of the protocol proposed by Bao and Yunger Halpern.}
    \label{fig_QMR1_Michael}
\end{figure}
In their seminal work, BYH~\cite{BaoHalpern2017} constructed a QMR constitution that is non-dictatorial yet satisfies quantum analogues of transitivity, unanimity, and independence of irrelevant alternatives, thereby violating a quantum version of Arrow's impossibility theorem. After introducing this quantum social welfare function, which they denote QMR, they turn to quantum voting tactics and strategies based on entanglement and interference, using a second constitution called QMR2. The latter does not satisfy Arrow's conditions and therefore does not itself violate Arrow's theorem. In the present work, QMR serves as the Arrow-violating benchmark constitution. Throughout the remainder of the paper we use the shorthand ``QMR'' for the ideal BYH QMR constitution and ``QMR2-inspired'' for our toy entanglement-based protocol, which follows the spirit of the QMR2 strategic scenarios but does not implement the full BYH QMR2 construction. 

Beyond the QMR constitution of BYH and the Schr\"odinger’s ballot framework of Sun et al.~\cite{Sun2021}, there is a small but growing body of work on quantum voting and quantum social choice. Some contributions focus on cryptographic and anonymity guarantees, proposing quantum anonymous voting schemes and optical or single-particle protocols that emphasize privacy and robustness against cheating or eavesdropping~\cite{HoroshkoKilin2009,Xu2022QuantumVotingNoMemory,Xiong2022SingleParticleVoting}. Other works pursue alternative formulations of quantum-computed voting rules and simplified majority mechanisms, for example via quantum logical operators or quantum-accelerated tallying algorithms that speed up winner determination under classical rules~\cite{QuantumLogicalVote2022,LiuHanXiaYu2023AcceleratingVoting}. Our study is complementary to these lines of research: rather than proposing a new cryptographic protocol or voting rule, we treat the BYH QMR constitution as a given and examine how its societal outcomes behave when instantiated, approximated, and stress-tested on noisy intermediate-scale quantum hardware.

We do not re-prove any Arrow-type result, nor do we claim that our noisy, finite-shot implementations satisfy the quantum unanimity or quantum IIA conditions exactly; instead, we take the ideal QMR constitution as given and use analytic and circuit-level experiments to study how realistic noise deforms its societal outcome distributions.

To employ quantum features, such as entanglement between voters, the QMR
constitution should be implemented on a real quantum computer, as suggested
in~\cite{BaoHalpern2017}. But if we are interested in just violating the quantum version of Arrow's theorem, we may do so by formulating a classical, quantum-inspired algorithm, borrowing only the conceptual mathematical structure of the original scheme. In such a scenario, the voters are asked to submit their ballots as sets of probabilities, as if their reduced and dephased states $\rho_i^d$ were measured and averaged to obtain the probabilities $p_i^{L}$'s. The classical part of QMR, which includes building the directed graph, running Tarjan's algorithm, and obtaining the SCCs order, is unchanged. Since the output from the previous step is in a single preference Hilbert space, $\rho_\text{soc}\in \mathcal{D}(\mathcal{H}_A)$, the \textit{give the minority a shot} and \textit{enforce unanimity} steps may be performed analytically, providing the resulting probabilities $\{p^L_\text{soc}\}$. The preference with the highest probability, $L_\text{soc}=\arg \,\underset{L}{\max}\, \{p^L_\text{soc}\}$, may be considered an election result. 

\subsection{This Work: Implementation and Stability of Quantum Voting}

This paper investigates the BYH QMR constitution via an analytic implementation complemented by circuit-level sampling experiments under realistic noise. The QMR constitution itself is evaluated classically: we formalize the preference basis, represent each voter’s ranking distribution, construct majority digraphs, identify strongly connected components via Tarjan’s algorithm~\cite{Tarjan1972}, and compute the resulting societal ranking distribution. In parallel, we implement the state-preparation and measurement stages as quantum circuits that encode voter states as superpositions over rankings and run them on noiseless simulators, noisy Qiskit FakeBackends~\cite{Qiskit}, and IBM superconducting hardware, using the devices as noisy samplers of the prescribed profile distributions. Comparing these empirical distributions to the analytic QMR baseline allows us to quantify how realistic noise deforms the societal outcome. To quantify stability, we compare ideal and noisy outcome distributions using Jensen–Shannon divergence and winner-flip rate. We evaluate two preference profiles that yield a practical assessment of when quantum voting retains its intended fairness properties on near-term hardware. 

In addition to implementing and stress-testing the QMR constitution, we also investigate a QMR2-inspired protocol that isolates the impact of entanglement between voters. This second component uses GHZ-type and separable superpositions over opposite rankings as a controlled test bed for how multi-voter quantum correlations affect majority outcomes, tie rates, and their degradation under simple noise channels.

At a high level, the main contributions of this work are:
\begin{itemize}
    \item We formalize an analytic implementation of the BYH QMR constitution that is directly applicable to probabilistic ranked ballots, including explicit constructions of majority digraphs, strongly connected components, and unanimity/minority-mixing operations.
    \item We present, to our knowledge, the first hardware-based stability study of QMR under realistic single-qubit and readout noise, combining noiseless simulators, Qiskit FakeBackends, and real IBM Quantum hardware, and quantify robustness using our own tailored metrics defined in Sec.~\ref{sub-section_metrics}.
    \item We introduce a QMR2-inspired entanglement-based toy model that uses GHZ-type voter blocks and mixed populations of separable and entangled voters as a controlled setting for studying how multi-voter quantum correlations affect winner frequencies and draw rates, and how fragile these effects are under simple local noise.
    \item We discuss layout and noise-design considerations for mapping analytic quantum voting constitutions to quantum circuits on noisy intermediate-scale quantum devices, highlighting trade-offs between qubit count, transparency of the preference encoding, and susceptibility to readout errors.
\end{itemize}

The remainder of the paper is structured as follows: Section \ref{section_methods} details the methods of both QMR and QMR2-inspired pipelines. Section \ref{section_results} presents experimental results and noise stability analysis for two representative QMR profiles and a QMR2-inspired entanglement protocol. Section \ref{section_conclusion_future_work} offers a discussion, concluding remarks, and future research directions.

\section{Methods}
\label{section_methods}
\subsection{General}

In this work, we implement both a classical baseline, the QMR quantum constitution, and a QMR2-inspired entangling variant, and we define metrics to compare (i) classical vs.\ quantum outcomes, (ii) noiseless vs.\ noisy quantum simulations, and (iii) simulated vs.\ real-hardware runs. The QMR pipeline follows the BYH constitution and is summarized in  Table~\ref{tab_QMR1_BYH}, which formalizes each numbered stage. The QMR2-inspired variant reuses the same preference-encoding and post-processing stack but augments the state-preparation stage with controlled entangling operations between selected voters, as detailed later in this section.

As outlined in Secs.~1.3--1.4, the QMR constitution is a genuinely quantum map
acting on a joint preference state $\sigma_{\mathrm{soc}}$. In the classical
ranked-voting literature, a Condorcet winner is defined, if any, as a candidate
that defeats every other candidate in pairwise majority comparisons. In the
present work, we evaluate the QMR map entirely by analytical
(classical) means: no quantum sampling is required to obtain the societal
distribution $\rho_{\mathrm{soc}}$ or to determine the Condorcet winner in this
sense. Quantum processing enters only when we choose to realize or estimate
the same profile distribution empirically (for example, via circuit-based
sampling on a simulator or quantum hardware).

\subsection{Classical voting scheme}
\label{Clasic_section}
To assess the functionality and stability of the quantum pipeline, we introduce a classical voting scheme that serves as a baseline for comparison. This scheme does not employ a majority digraph or strongly connected components. Instead, the societal winner is determined directly from the expected pairwise preferences. Concretely, for each ordered pair $(a,b)$ we compute the expected fraction of voters preferring $a$ to $b$, and the classical Condorcet winner (if it exists) is the candidate that beats every other alternative in these expected pairwise comparisons, as shown in these steps:
\begin{enumerate}
    \item \textbf{Pairwise expectations.}
          For each ordered pair $(a,b)$ and voter $v$, let $p_v(a\succ b)$ be the probability (under $v$'s distribution) that $a$ ranks above $b$. Define the expected number of voters preferring $a$ to $b$,
         \begin{equation}
             E_{a\succ b}\;=\;\sum_{v=1}^n p_v(a\succ b),\qquad
              E_{b\succ a}\;=\;\sum_{v=1}^n p_v(b\succ a).
         \end{equation}

    \item \textbf{Pairwise winners.}
          For each pair $(a,b)$, if $E_{a\succ b}>E_{b\succ a}$ we say $a$ beats $b$.
    \item \textbf{Condorcet check.}
          A candidate $c$ is the (probabilistic) Condorcet winner if $c$ beats all other candidates pairwise (wins $m\!-\!1$ pairwise contests for $m$ candidates).
\end{enumerate}
This generalizes the Condorcet principle to non-deterministic ballots using only arithmetic on input distributions.

\subsection{QMR – The main quantum voting scheme}
\label{subsec_QMR_main}

In this work, the BYH QMR constitution is implemented as a deterministic algorithm on classical probability distributions over rankings.  
Each voter supplies a probability vector over strict rankings; for every realized profile, the constitution constructs an $\varepsilon$-aware majority digraph, computes the strongly connected components (SCCs), and assigns a profile-conditioned distribution over rankings.  
These per-profile distributions are then aggregated into a single societal distribution $\rho_{\text{soc}}$ over rankings.  
Winner identification (Condorcet winner or top set) and any optional sampling of a concrete ranking are performed from $\rho_{\text{soc}}$ or from its empirical estimator obtained from simulator or hardware runs.  
All steps up to this final sampling are computed analytically and exactly.

We adopt notation aligned with Table~\ref{tab_QMR1_BYH}.  
Objects written as $\rho_i$, 
and $\rho_{\text{soc}}$ in the original quantum formulation are implemented as classical probability vectors.  
We denote by $\chi^{(1)}(\vec{L})$ the base per-profile distribution over rankings obtained from SCCs and linear extensions, and by $\chi_{\text{soc}}(\vec{L})$ the refined per-profile distribution after applying the ``give the minority a shot'' (GMS) step and unanimity enforcement.  
The societal distribution is then given by
\[
  \rho_{\text{soc}} \;=\; \sum_{\vec{L}} p(\vec{L})\,\chi_{\text{soc}}(\vec{L}),
\]
which in the code is represented as a probability vector over $\mathcal{L}(A)$.

{
\subsubsection*{Algorithm steps}
\begin{enumerate}
    \item As input, the algorithm receives the voters' preference probabilities\newline 
    $$\{p_i^{L_i}\}=
    \left(p_1^{L_{1,(1)}},\ldots,p_1^{L_{1,(M)}},\ldots,
    p_n^{L_{n,(1)}},\ldots,p_n^{L_{n,(M)}}\right),$$ where $i\in V$, $L_{i,(j)}\in\mathcal{L}(A)$, and $M=|\mathcal{L}(A)|=m!$.
    \item The following steps: (a) construction of a directed graph, (b) obtaining the list of \textit{strongly connected components} using Tarjan's algorithm, and (c) formation of linear extensions, are as described in~\cite{BaoHalpern2017}. The output is the quantum state $\chi^{(1)}(\vec{L})$, formulateddescribed analytically, and used as such in the following steps.
    \item The GMS and EU steps in~\cite{BaoHalpern2017}, acting on $\chi^{(1)}(\vec{L})$ and taking $\{p_i^{L_i}\}$ as input, produce the societal preference state, $\rho_\text{soc}$, in a classical computation. 
    \item The resulting state, $\rho_\text{soc}$, obtained analytically, is prepared as a quantum state (either in simulation or on real quantum hardware) and then measured in the preference basis, $B$, providing the election outcome.
\end{enumerate}
}

\subsubsection*{Shot-based emulation on simulator, fake backends, and hardware}

Conceptually, the QMR constitution is a premeasurement map defined entirely at the level of probability distributions over rankings.  
In the experiments, this analytically defined distribution $\rho_{\mathrm{soc}}$ is emulated on quantum simulators, IBM fake backends, and real quantum hardware by running circuits whose measurement statistics yield an empirical estimator $\widehat{\rho}_{\mathrm{soc}}$.  
The discrepancy between $\widehat{\rho}_{\mathrm{soc}}$ and the analytically computed $\rho_{\mathrm{soc}}$ is quantified using the Jensen–Shannon divergence metric $\widehat{JS}_{\mathrm{div}}\bigl(\widehat{\rho}_{\mathrm{soc}},\rho_{\mathrm{soc}}\bigr)$, with $\rho_{\mathrm{soc}}$ serving as the noiseless baseline.

\subsubsection{Quantum noise settings}

All simulations are implemented in Qiskit~\cite{Qiskit} using the \texttt{AerSimulator} backend, with explicit \texttt{NoiseModel} and \texttt{ReadoutError} channels to model readout noise where specified.

To examine the robustness of QMR under realistic experimental conditions, we implemented the voting circuits on three types of execution backends: a noiseless simulator, noisy simulators, and real IBM Quantum devices. The noiseless case uses Qiskit’s \texttt{AerSimulator} without any attached noise model and serves as the reference baseline with which all noisy runs are compared.

Noisy simulations are carried out using \texttt{AerSimulator} equipped with explicit \texttt{NoiseModel} objects from \texttt{qiskit}. In this work we use standard single-qubit channel models (e.g.\ depolarizing, bit-flip, and phase-flip errors) and a classical readout-error channel implemented via \texttt{ReadoutError}. Each channel is parameterised by a probability $p\in[0,1]$ that sets the strength of the corresponding noise process. For the stability sweeps reported here, we focus on readout-dominated regimes: gate noise is disabled and only the readout-error probability is varied across a range of amplitudes, while all other aspects of the circuit and compilation are kept fixed. 

Device-level noise is probed using Qiskit FakeBackend models (such as \texttt{FakeBrisbane}) and real IBM Quantum hardware (in particular the \texttt{ibm\_torino} device). In these runs, we do not impose any additional handcrafted noise model. Instead, we rely on the native calibration data and noise characteristics of the backend. In all simulator, FakeBackend, and hardware experiments, a “shot” corresponds to one full circuit execution, and empirical outcome distributions are obtained by repeating the circuit for a fixed number of shots per setting.

\medskip

\subsubsection{Metrics}
\label{sub-section_metrics}
To quantify the stability and robustness of the QMR pipeline, we define three
primary outcome-level metrics---winner-agreement rate
$\Gamma_{\mathrm{win}}$, normalized Condorcet-winner flip rate
$\tilde{\gamma}$, and Jensen--Shannon divergence
$\widehat{JS}_{\mathrm{div}}$. These quantities are deliberately
complementary: $\Gamma_{\mathrm{win}}$ captures how often the quantum pipeline
reproduces a chosen benchmark winner, $\tilde{\gamma}$ measures the temporal
stability of that winner across repeated runs, and
$\widehat{JS}_{\mathrm{div}}$ probes the full deformation of the societal
distribution over rankings. 

For each QMR experiment and execution mode, we therefore report these three metrics jointly, which allows us to identify noise thresholds at which QMR departs from its classical benchmark, distinguish benign distributional drift from real winner instability, and systematically compare simulator, FakeBackend, and hardware behavior on a common footing.

\vspace{0.6em}
\noindent\textbf{1.\ Winner–Agreement Rate $\Gamma_{\mathrm{win}}$.}
The first and most interpretable metric is the winner-agreement rate. Operationally, it quantifies the frequency with which the quantum pipeline returns the same Condorcet winner as a chosen reference scheme.

In our case, the reference is the classical pipeline applied once to the voters’ distributions, which yields either a single classical Condorcet winner or a declaration of ``no unique Condorcet winner''. We denote this reference outcome by $CW_{\mathrm{classical}}$. For each quantum run $i$ (noiseless simulator, noisy simulator, FakeBackend, or hardware), we denote by $CW_i$ the Condorcet winner inferred from the corresponding QMR societal distribution, or set $CW_i := \varnothing$ if that run has no unique Condorcet winner.

If the classical pipeline yields no unique Condorcet winner, we set
\(CW_{\mathrm{classical}} := \varnothing\) and evaluate the same formula, so runs are counted precisely when \(CW_i = \varnothing\). We then define $\Gamma_{\mathrm{win}}$ as the fraction of runs where the quantum Condorcet winner $CW_i$ coincides with the classical reference $CW_{\mathrm{classical}}$:
\begin{equation}
\Gamma_{\mathrm{win}}
=\frac{1}{N}\sum_{i=1}^{N}\mathbf{1}\!\bigl\{\,CW_i = CW_{\mathrm{classical}}\,\bigr\}.
\label{eq:win-agree}
\end{equation}

We use $\Gamma_{\mathrm{win}}$ as a benchmark-consistency metric: when
$\Gamma_{\mathrm{win}}\approx 1$, the quantum scheme effectively inherits the
classical decision for the given profile; when $\Gamma_{\mathrm{win}}\approx 0$,
the quantum constitution has settled into a markedly different majority outcome.
This makes $\Gamma_{\mathrm{win}}$ a natural way to identify noise thresholds
beyond which hardware or noise-model runs can no longer be regarded as
faithfully reproducing the classical benchmark, even if they remain internally
stable.

\vspace{0.6em}
\noindent\textbf{2.\ Normalized CW Flip Rate $\tilde{\gamma}$.}
Winner agreement by itself does not reveal how stable the quantum winner
is across repeated runs at fixed settings. A scheme may agree with the classical
winner on average, yet fluctuate noticeably from run to run. To capture this
temporal stability, we define the normalized Condorcet-winner flip rate
$\tilde{\gamma}$.

For a batch of $N$ runs with winners $CW_i$ and baseline winners
$CW^{(0)}_i$ (from a noiseless simulator with the same circuit and shot
count), we first compute
\begin{equation}
\gamma_{\text{run}}=\frac{1}{N-1}\sum_{i=1}^{N-1}\mathbf{1}\!\bigl\{CW_i\neq CW_{i+1}\bigr\},\qquad
\gamma_{0}=\frac{1}{N-1}\sum_{i=1}^{N-1}\mathbf{1}\!\bigl\{CW^{(0)}_i\neq CW^{(0)}_{i+1}\bigr\},
\label{eq:fliprate-components}
\end{equation}
which are the raw flip rates between adjacent runs for the noisy and
noiseless settings, respectively. With a pseudo-count
\(\epsilon_{\mathrm{pc}}=\tfrac{1}{N}\) for numerical stability, we then define
\begin{equation}
\tilde{\gamma}=
\begin{cases}
\dfrac{\gamma_{\text{run}}}{\epsilon_{\mathrm{pc}}}, & \text{if }\gamma_{0}=0,\\[0.8em]
\dfrac{\gamma_{\text{run}}}{\gamma_{0}+\epsilon_{\mathrm{pc}}}, & \text{otherwise.}
\end{cases}
\label{eq:fliprate}
\end{equation}

Thus, $\tilde{\gamma}\approx 1$ indicates that the noisy configuration is about
as stable as its own noiseless baseline, while $\tilde{\gamma}\gg 1$ signals
strongly enhanced volatility in the winner. This normalization is important for
two reasons. First, it corrects for any residual flip rate that may arise
purely from finite-shot sampling in the baseline. Second, it makes flip rates
comparable across different batch sizes and experiments: a large increase in
$\tilde{\gamma}$ directly reflects a degradation in the reproducibility of the
Condorcet decision. In this sense, $\tilde{\gamma}$ functions as a
winner-stability metric that is complementary to the agreement rate
$\Gamma_{\mathrm{win}}$.

\vspace{0.6em}
\noindent\textbf{3.\ Jensen--Shannon Divergence $\widehat{JS}_{\mathrm{div}}$.}
Both $\Gamma_{\mathrm{win}}$ and $\tilde{\gamma}$ operate only on the winner label (or its absence) extracted from each run, rather than on the full societal ranking distribution.
However, the QMR
constitution is defined at the level of an entire societal distribution over
rankings $\rho_{\mathrm{soc}}$, which can change substantially even when the
winner remains the same. To probe these finer deformations, we track the
Jensen--Shannon divergence between the induced societal distributions in noisy
runs and the corresponding noiseless baseline.

Concretely, we define
\begin{equation}
\widehat{JS}_{\mathrm{div}}
=\frac{1}{N}\sum_{i=1}^{N}\operatorname{JS}\!\bigl(\rho^{(\text{run})}_i \,\Vert\, \rho^{(0)}\bigr),
\label{eq:jsdiv}
\end{equation}
where $\operatorname{JS}$ uses base-2 logarithms so that
$\operatorname{JS}\in[0,1]$. Here $\rho^{(\text{run})}_i$ is the run-$i$
induced distribution over rankings
\[
\text{(profile)} \;\to\; \text{(majority digraph)} \;\to\; \text{(SCCs)} \;\to\;
\text{(linear extensions)} \;\to\; \text{(aggregation)},
\]
and $\rho^{(0)}$ is the noiseless-simulator counterpart under the same circuit and number of shots. We choose the Jensen--Shannon divergence rather than, for
example, the Kullback--Leibler divergence because JS is symmetric, always
finite, and bounded between $0$ and $1$, enabling direct visual comparison
across experiments and noise levels.

In the QMR analysis framework, $\widehat{JS}_{\mathrm{div}}$ serves as a
distribution-level deformation metric: small values indicate that the
shape of the societal distribution over rankings is close to its ideal form,
even if some finite-shot randomness affects individual runs, while large values
signal that noise has qualitatively reshaped the collective preferences,
typically by flattening or broadening the distribution. Importantly,
$\widehat{JS}_{\mathrm{div}}$ can be large even when $\Gamma_{\mathrm{win}}$
remains close to $1$, revealing regimes where the winner is stable but the
underlying ranking probabilities have drifted significantly away from the ideal
quantum constitution.

\vspace{0.6em}
Conceptually, the three primary metrics are kept distinct to separate benchmark agreement, temporal stability, and full-distribution deformation. This separation is important for interpreting
our results of QMR experiments. Taken together, these metrics allow us to distinguish regimes in which QMR either closely reproduces the classical Condorcet outcome (as in Exp.~2 at low to moderate readout noise), or remains winner-stable while drifting to a reproducible non-classical equilibrium that departs from a classically Condorcet-consistent benchmark (as in Exp.~1 once readout noise exceeds a modest threshold), or exhibits both large distributional deviations and frequent winner flips under extreme noise, particularly on device-like backends.
.

\subsection{QMR2-inspired protocol: entanglement effect}
\label{subsec:QMR2_methods}

In addition to QMR, we also investigate a QMR2-inspired constitution, closely following the BYH~\cite{BaoHalpern2017} QMR2 protocol, as a testbed for assessing the impact of entanglement between voters. Whereas QMR operates on dephased, product-structured profiles and reconstructs a societal ranking via majority digraphs and SCC structure (Sec.~\ref{section_methods}), QMR2-inspired protocol is designed to highlight how joint quantum correlations in the preference state of a group of voters can affect the outcome, even when the aggregation rule is relatively simple.

We consider a subset of $k$ voters who may either (i) hold separable superpositions over two opposite rankings, or (ii) share a genuinely entangled state over the same pair of rankings. Following BYH~\cite{BaoHalpern2017}, let $L$ denote a linear order and $\bar{L}$ its reversed (anti-cycle) order\footnote{By anti-cycle here we mean that for $L=(a_1,a_2,\ldots,a_{m-1},a_m)$, we define $\bar{L}=(a_m,a_{m-1},\ldots,a_2,a_1)$} . For a single voter, a superposition of these two preferences can be written as
\begin{equation}
\ket{\psi_i} 
  \;=\; c_{i,L}\ket{L} + c_{i,\bar{L}}\ket{\bar{L}},
\end{equation}
with $\lvert c_{i,L}\rvert^2 + \lvert c_{i,\bar{L}}\rvert^2 = 1$.  

For a block of $k$ voters, we contrast two following idealized scenarios:\\

1. \textbf{Entangled (GHZ-type) preference block.}  
   All $k$ voters are jointly prepared in a generalized GHZ state over $L$ and $\bar{L}$,
   \begin{equation}
       \ket{\psi_{\mathrm{GHZ}}}
       \;=\;
       \frac{1}{\sqrt{2}}\Bigl(\ket{L}^{\otimes k} + \ket{\bar{L}}^{\otimes k}\Bigr),
       \label{Eq:GHZ}
   \end{equation}
   so that their preferences are perfectly correlated: namely, all the $k$ voters in a subgroup either choose $L$ or $\bar{L}$.

2. \textbf{Separable preference block.}  
   Each of the $k$ voters independently holds a local superposition of the same two rankings,
   \begin{equation}
       \ket{\psi_{\mathrm{sep}}}
       \;=\;
       2^{-k/2}\Bigl(\ket{L} + \ket{\bar{L}}\Bigr)^{\otimes k},
       \label{eq:sep}
   \end{equation}
   so that, in each run, some subset of the voters may realize $L$ and the rest $\bar{L}$, with binomially distributed counts.

Both constructions share the same single-voter reduced density matrix and thus the same local statistics. The distinction is purely in the multi-voter correlations: in the GHZ case, all voters always agree; in the separable case, they may disagree from run to run. QMR2 is therefore a natural setting to ask how such correlation structure affects the distribution of majority outcomes and the probability of ties.

Operationally, our simulation proceeds as follows:
\begin{enumerate}
    \item Fix the number of candidates (here $m=3$, so there are $3!=6$ possible strict rankings) and a subset of size $k$ (here $k=5$ voters per group).
    \item For each of the $10{,}000$ iterations, prepare either $\ket{\psi_{\mathrm{GHZ}}}$ or $\ket{\psi_{\mathrm{sep}}}$ for the $k$ voters, measure all voters in the preference basis, and record the resulting classical rankings $(L_1,\dots,L_k)$.
    \item Apply a simple majority rule within the group: the group outcome is the ranking (or candidate) that appears most frequently in that iteration. If there is no unique majority (e.g.\ a tie between two or more rankings), we record a draw outcome, denoted by $-1$.
    \item Repeat this procedure for:
          (i) fully matched groups (all $k$ voters in the GHZ or separable block), and  
          (ii) mixed groups where a fixed fraction of the voters are drawn from an independent “random” population with uniform preferences over the rankings.
\end{enumerate}
The resulting empirical frequencies are then compared between the GHZ and separable scenarios under identical conditions, allowing us to isolate the effect of entanglement on the rate of conclusive majority decisions and on the expectation values of a preference observable.
The empirical frequencies obtained in this setup are reported in
Sec.~\ref{Results_QMR2}, where we compare GHZ-type and separable
blocks under identical conditions and under local bit-flip noise.

\subsubsection{Preference basis map}
\label{subsubsec:preference_basis_map}

In our implementation, a quantum register consisting of $q$ qubits is mapped into the preference basis using a Lehmer-code representation. Since we exclude scenarios in which a ranking contains a tie, the set of all strict rankings forms a cyclic structure in both the qubit representation and the induced preference basis. The transformation
\begin{equation}
    \mathcal{E}\!\left(
        \sum_{i_0,\dots,i_{q-1}}
        \psi_{i_0,\dots,i_{q-1}}
        \ket{i_0,\dots,i_{q-1}}
    \right)
    =
    \sum_{L \in \mathcal{L}(A)}
        \psi_{L}\ket{L}
        \label{eq:Psi_gamma_ket}
\end{equation}
is defined using the Lehmer index of each computational-basis state that is assigned to a valid ranking. Because
\[
    2^{q} \;\ge\; m!,
\]
the map necessarily involves redundant labels: only $m!$ out of the $2^{q}$ computational-basis states are used to represent legitimate strict rankings, while the remaining $2^{q}-m!$ labels do not encode any preference and are excluded by construction.

At the Hilbert-space level, we therefore distinguish between the full physical space
\[
    \mathcal{H}_{\mathrm{phys}} = (\mathcal{H}_2)^{\otimes q},
\]
and the valid subspace $\mathcal{H}_{\mathrm{phys}}^{\mathrm{valid}}\subset \mathcal{H}_{\mathrm{phys}}$ spanned by the $m!$ computational-basis states that are actually used for rankings. The preference Hilbert space $\mathcal{H}_{A}$ is defined as
\[
    \mathcal{H}_{A} := \mathrm{span}\{\ket{L}:L\in\mathcal{L}(A)\},
    \qquad
    \dim(\mathcal{H}_{A}) = m!,
\]
and the encoding map is
\[
    \varepsilon : \mathcal{H}_{\mathrm{phys}}^{\mathrm{valid}} \longrightarrow \mathcal{H}_{A},
    \qquad
    \dim(\mathcal{H}_{\mathrm{phys}}^{\mathrm{valid}}) = \dim(\mathcal{H}_{A}) = m!.
\]
Restricted to $\mathcal{H}_{\mathrm{phys}}^{\mathrm{valid}}$, $\mathcal{E}$ is a bijection: every valid physical label corresponds to exactly one strict ranking, and vice versa. The remaining basis states in $\mathcal{H}_{\mathrm{phys}}$ represent redundant, non-physical labels from the voting perspective and are never populated in our state-preparation routines.

For entangled states, the Lehmer representation operates identically: each computational-basis component in the entangled superposition that lies in $\mathcal{H}_{\mathrm{phys}}^{\mathrm{valid}}$ has a well-defined Lehmer index, so every valid quantum state admits a corresponding representation in the preference basis. In multi-party scenarios, one may enlarge the encoded Hilbert space to $\mathcal{H}_{A'}$ of dimension $2^{kq}$ when $k$ voters are encoded, and the redundancy of unused labels scales accordingly.

Measurements are carried out locally in the computational basis using standard PVMs, with eigenvalues $\ket{0}\mapsto +1$ and $\ket{1}\mapsto -1$. No multi-qubit correlation measurements or high-order $t$-designs are required: the Lehmer encoding ensures that all the relevant correlation structure is already reflected in the computational-basis outcomes. Local projective measurements followed by classical decoding via the Lehmer map suffice to reconstruct the joint ranking distribution used in the QMR2 majority-rule analysis.

\section{Results}

\label{section_results}
We present two classes of experiments. 
First, we study the robustness of the QMR constitution under readout and device-level noise for two representative five-voter profiles (Exp.~1 and Exp.~2). 
Second, we analyze a QMR2-inspired entanglement-based protocol to quantify how multi-voter quantum correlations affect the rate and stability of majority outcomes, both in ideal settings and under a simple local bit-flip noise model.
We thus report results for both the QMR constitution and the QMR2-inspired protocol.

At a high level, the QMR experiments map out a clear noise threshold: for small readout amplitudes, the quantum outcomes closely track the classical scheme, while beyond a modest noise leve,l they converge to structured but non-classical societal distributions. The QMR2-inspired simulations, in turn, show that entangled voter blocks can eliminate draw outcomes in ideal conditions, yet this advantage is fragile under local bit-flip noise and degrades toward effectively random behavior.

In QMR, we show the results for two representative five-voter experiments—denoted Exp.~1 and Exp.~2—each implemented under five execution modes: the classical voting scheme, a noiseless quantum simulator, a noisy quantum simulator, an IBM FakeBackend (e.g., FakeBrisbane, FakeTorino), and real IBM Quantum hardware. Each mode instantiates the QMR constitution described in Sec.~\ref{section_methods}, except where noted. Winner–agreement $\Gamma_{\mathrm{win}}$ is evaluated against the classical scheme, while the divergence $\widehat{JS}_{\mathrm{div}}$ and flip rate $\tilde{\gamma}$ are referenced to the noiseless simulator baseline for the same circuit and shot count. 

Throughout the QMR experimen,ts we fix $m=3$ candidates and $n=5$ voters. This choice reflects the smallest non-trivial setting in which Condorcet cycles and non-transitive majority digraphs can occur ($m \ge 3$), while keeping the profile space ($6^5$ joint rankings) and SCC-based mixing numerically tractable. At the circuit level, $m=3$ and $n=5$ match the 15-qubit per-voter encoding introduced in Sec.~\ref{section_methods}, which can be mapped to currently available IBM devices with enough remaining qubits to accommodate routing and simple multi-voter interactions. Increasing the electorate or candidate set (for example to $n=7$, $m=4$) would simultaneously enlarge the classical state space and require more physical qubits or deeper reuse schemes, making exhaustive noise sweeps and repeated hardware runs prohibitively costly on present-day hardware.

Conceptually, we distinguish four independent error channels—depolarizing $(\eta_{d})$, bit-flip $(\eta_{\mathrm{bf}})$, phase-flip $(\eta_{\mathrm{ph}})$, and readout $(\eta_{\mathrm{ro}})$—with nominal error probabilities $p\in[0.01,0.5]$. In our initialize-measure circuits with computational-basis readout, pure phase-flip noise acting on already diagonal states does not change outcome probabilities. In practice, moderate gate-level depolarizing and Pauli errors produced only marginal changes in the societal distributions in our numerical tests. By contrast, readout error (and, where relevant, bit-flip) had a clear and systematic effect on winner frequencies and ranking distributions. For this reason, our explicit noise sweeps and reported metrics focus on $\eta_{\mathrm{ro}}$ (and $\eta_{\mathrm{bf}}$), treating depolarizing and phase-flip channels as subdominant in the regimes studied.

\begin{figure}[H]
\centering
\includegraphics[width=0.7\textwidth]{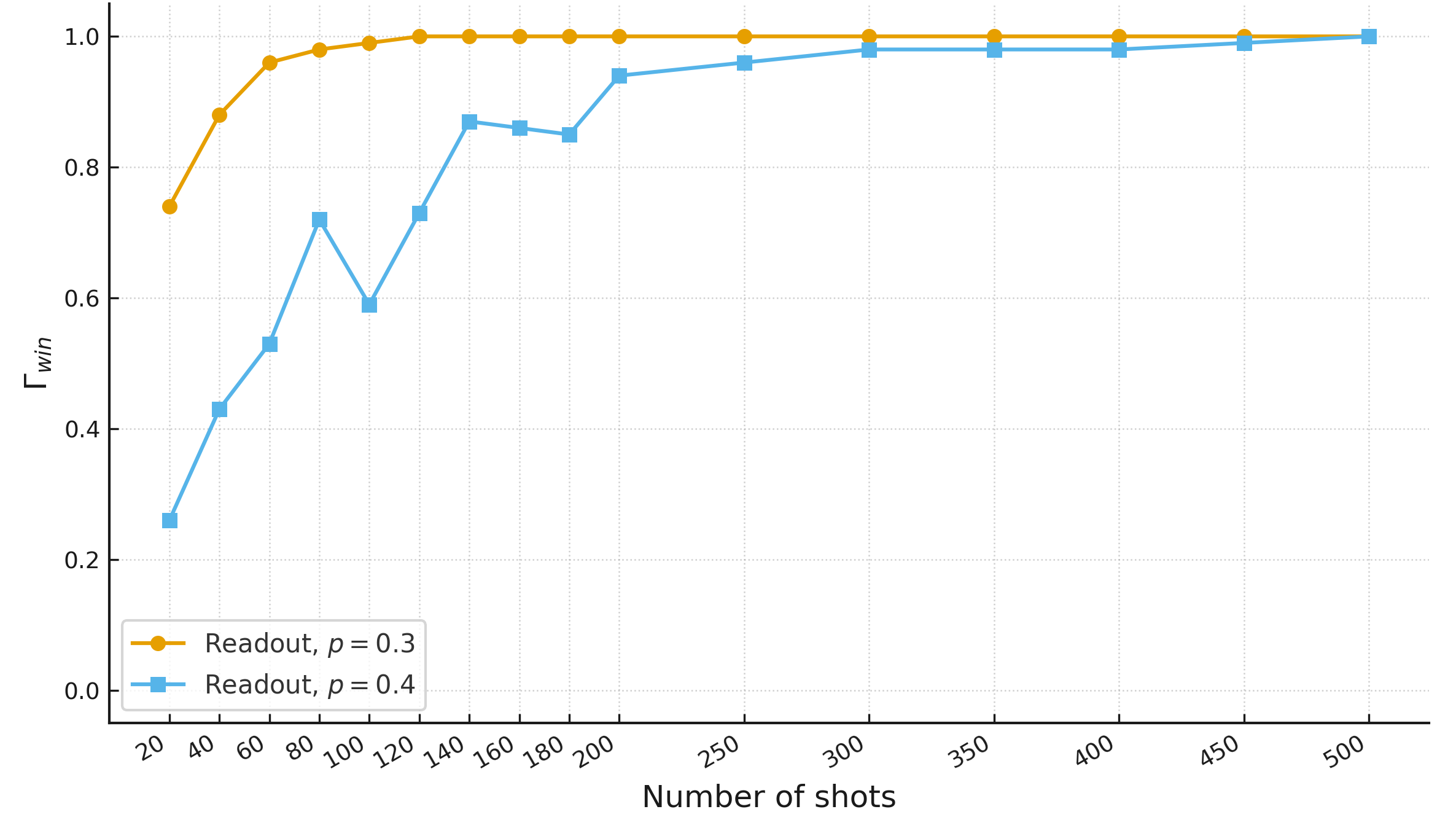}
\caption{{\bf Winner agreement convergence.} The winner agreement metrics convergence as a function of the number of shots in experiment 1, per readout noise amplitude.}
\label{fig:win_agreement_per_shots}
\end{figure}

Shot budgets and batch sizes were chosen to balance statistical precision with resource constraints. For all noise-model and FakeBackend simulation,s we used \(500\) shots per run and \(100\) repetitions per configuration, which reduces sampling variability and stabilizes Condorcet decisions and Jensen--Shannon divergence estimates. In Figure~\ref{fig:win_agreement_per_shots}, one can see that the slopes of the winner–agreement curves change as a function of readout noise. High-shot runs (small $1/\sqrt{S}$ sampling noise) reveal a sharp drop in $\Gamma_{\mathrm{win}}$ as readout noise shifts the true societal distribution away from the classical Condorcet winner, whereas low-shot runs add enough sampling noise to blur this transition and keep $\Gamma_{\mathrm{win}}$ artificially high over a wider noise range.
Runs on an IBM FakeBackend were included to approximate real-hardware behavior as closely as possible while remaining reproducible, capturing device-like routing on an actual coupling map, native-gate characteristics, and realistic readout noise. Hardware runs were limited to \(50\) shots per run and \(10\) repetitions (500 shots total) by queuing and access constraints. At this shot count, the standard error on an individual winner frequency is on the order of \(0.02\) for probabilities near \(1/2\), which is small relative to the differences induced by the noise levels studied, so winner–agreement and flip-rate estimates remain meaningful despite the more modest statistics. Hardware results are reported alongside the corresponding simulator and FakeBackend baselines in the figures below. For every batch, we logged the Condorcet winner, the number and sizes of strongly connected components (SCCs) of the majority digraph, and the Jensen--Shannon divergence \(\widehat{JS}_{\mathrm{div}}\) against the corresponding noiseless baseline; from these logs we then computed the summary metrics \(\Gamma_{\mathrm{win}}\), \(\tilde{\gamma}\), and \(\widehat{JS}_{\mathrm{div}}\) plotted in the following subsections.

\subsection{QMR Experiment 1: Stable Non-Agreement Profile}
\subsubsection{Experiment profile}
Voters are divided as follows: the first voter assigns full weight to ranking~1 (ACB), the second to ranking~2 (BAC), the third to ranking~3 (BCA), the fourth voter to ranking~4 (CAB), and the last voter to ranking~5 (CBA). Thus, five distinct rankings are populated out of the six possible strict orders on $\{A,B,C\}$.

This profile admits a classical Condorcet winner. A direct pairwise tally gives
$N_{BA}=3>2$, $N_{CA}=3>2$, and $N_{CB}=3>2$, so candidate~$C$ defeats both $A$ and $B$ in majority comparisons. The profile is therefore Condorcet-consistent but far from unanimous: each candidate receives non-negligible support, and the majority margins are relatively small.

\begin{figure}[H]
\centering
\includegraphics[width=0.7\textwidth]{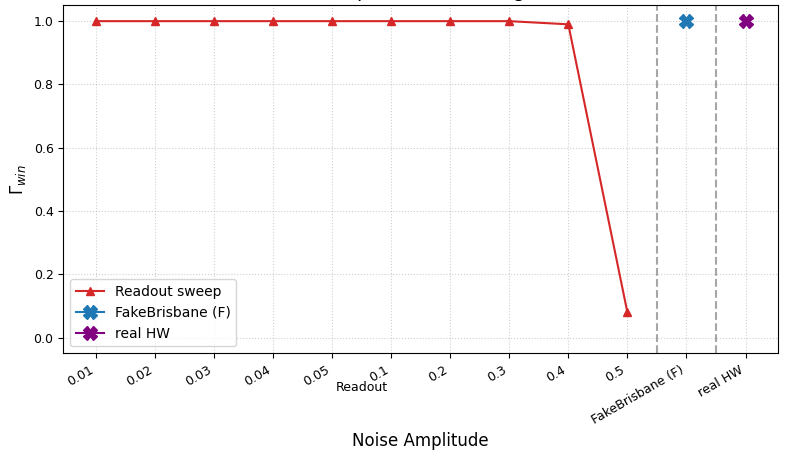}
\caption{{\bf Experiment~1 -- Winner–agreement rate. Plot of} \(\Gamma_{\mathrm{win}}\) under readout noise.}
\label{fig:exp1_readout_only_winner_agreement}
\end{figure}

\begin{figure}[H]
\centering
\includegraphics[width=0.7\textwidth]{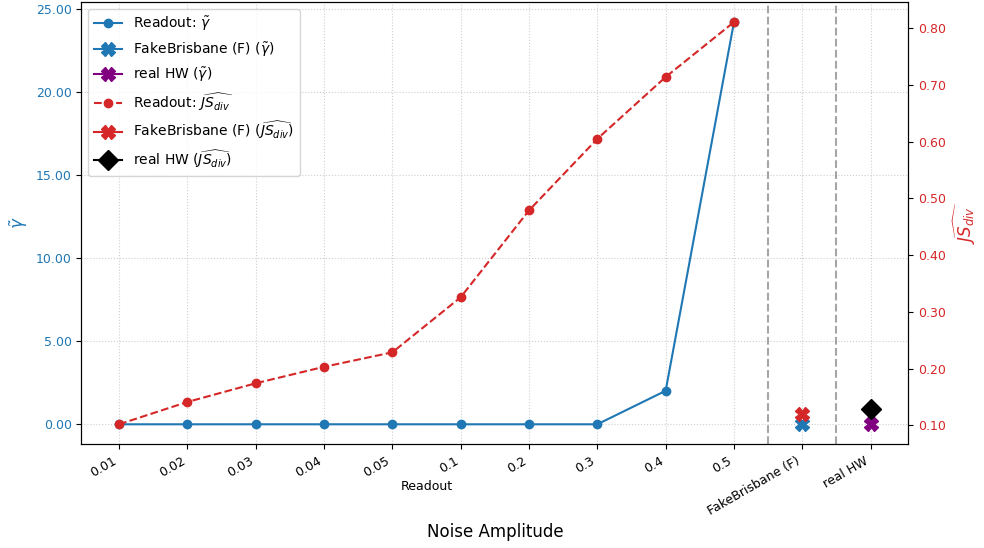}
\caption{{\bf Experiment~1 -- Divergence.} Plot of $\widehat{JS}_{\mathrm{div}}$ and the flip-rate $\tilde{\gamma}$ under readout noise.}
\label{fig:exp1_readout_only_2Metrics}
\end{figure}

Because the classical Condorcet winner is unambiguous yet backed only by relatively narrow pairwise majorities, this configuration is useful as a ``stable non-agreement’’ test case: at low readout noise the QMR pipeline tracks the classical Condorcet winner $C$, whereas above a noise threshold of 0.4, the societal distribution drifts toward a reproducible non-classical equilibrium that disagrees with the classical benchmark, as quantified in Figs.~\ref{fig:exp1_readout_only_winner_agreement}--\ref{fig:exp1_readout_only_2Metrics}.

\subsubsection{Observed behavior}
Figures~\ref{fig:exp1_readout_only_winner_agreement} and~\ref{fig:exp1_readout_only_2Metrics} show a smooth, monotonic increase of $\widehat{JS}_{\mathrm{div}}$ with the readout-noise amplitude, from $\widehat{JS}_{\mathrm{div}}\approx0.10$ at $p=0.01$ to $\widehat{JS}_{\mathrm{div}}\approx0.71$ at $p=0.4$, and $\widehat{JS}_{\mathrm{div}}\approx0.81$ at $p=0.5$.  
Throughout the range $0.01 \le p \le 0.4$ the normalized flip rate remains zero ($\tilde{\gamma}=0$) and the winner–agreement stays perfect ($\Gamma_{\mathrm{win}}=1$), indicating that the Condorcet winner is completely stable even though the societal distribution gradually drifts away from the noiseless reference.  
Only at the largest readout amplitude, $p=0.5$, do we observe a sharp transition: the flip rate jumps to $\tilde{\gamma}\approx 6.1$ and the agreement collapses to $\Gamma_{\mathrm{win}}\approx 0.02$.

Thus, for a wide interval of readout amplitudes, the system exhibits a gradual redistribution of probability mass without any volatility in the winning set, followed by an abrupt breakdown into an SCC-dominated regime only at very high noise.

The device-level points fall squarely in the robust regime.  
Both the FakeBrisbane backend and the real IBM hardware exhibit $\Gamma_{\mathrm{win}}=1$, $\tilde{\gamma}=0$, and modest divergences $\widehat{JS}_{\mathrm{div}}\approx 0.12$--$0.13$, comparable to synthetic readout noise at $p\approx 0.02$–$0.03$.  
In other words, realistic device noise for this experiment deforms the societal distribution slightly but does not alter the Condorcet winner or the acyclicity of the majority digraph.

\medskip
\noindent\textbf{Summary.}
Agreement with the classical Condorcet winner remains perfect up to $p\approx 0.4$, while $\widehat{JS}_{\mathrm{div}}$ rises smoothly towards $\approx 0.8$ and the SCC structure gradually compresses.  
Only at the extreme amplitude $p=0.5$ do we observe simultaneously large divergence, high flip rate, and near-zero winner–agreement, signaling a qualitative change in the majority structure towards a large top cycle.  
The FakeBackend and real-hardware points lie well below this threshold, underscoring that, for Experiment~1, the QMR constitution is highly robust to realistic levels of readout and device noise.

\bigskip

\subsection{QMR Experiment 2: Condorcet Profile with Wide Pairwise Majorities}
\subsubsection{Experiment profile}
Voters are divided into three blocks: two voters assign full weight to ranking~0 (ABC), two to ranking~1 (ACB), and one to ranking~2 (BAC). This yields a \(2\text{–}2\text{–}1\) profile in which candidate~A receives decisive support: A is ranked first by four of the five voters and is the unique classical Condorcet winner. A direct pairwise tally gives
\[
N_{AB}=4>1,\qquad N_{AC}=5>0,\qquad N_{BC}=3>2,
\]
so the majority relation forms the strict total order
\[
A \succ_{\mathrm{maj}} B \succ_{\mathrm{maj}} C.
\]
The corresponding majority digraph has three singleton strongly connected components \(\{A\},\{B\},\{C\}\); its SCC structure is therefore robust in the sense that moderate perturbations must flip multiple pairwise relations before altering the Condorcet winner or the top SCC. In particular, candidate~A defeats \(B\) and \(C\) by wide pairwise margins (4--1 and 5--0, respectively), in contrast to the narrow 3--2 margins of Exp.~1.

\begin{figure}[H]
\centering
\includegraphics[width=0.7\textwidth]{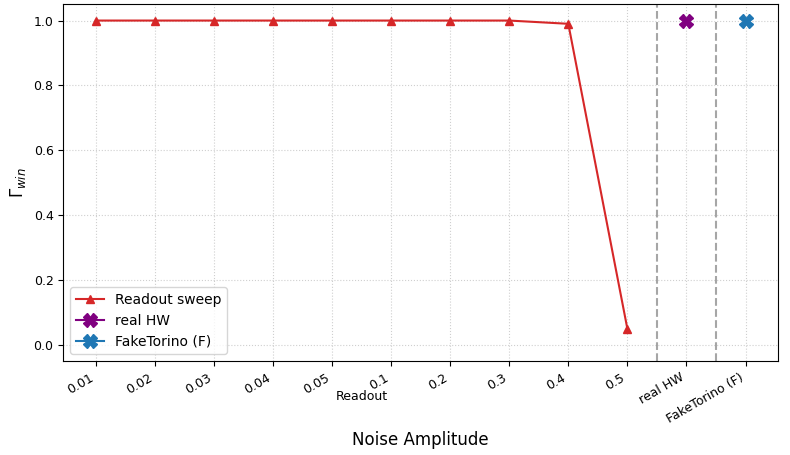}
\caption{{\bf Experiment~2 -- Winner–agreement rate.} Plot of \(\Gamma_{\mathrm{win}}\) under readout noise.}
\label{fig:exp2_winner_agreement}
\end{figure}

\begin{figure}[H]
\centering
\includegraphics[width=0.7\textwidth]{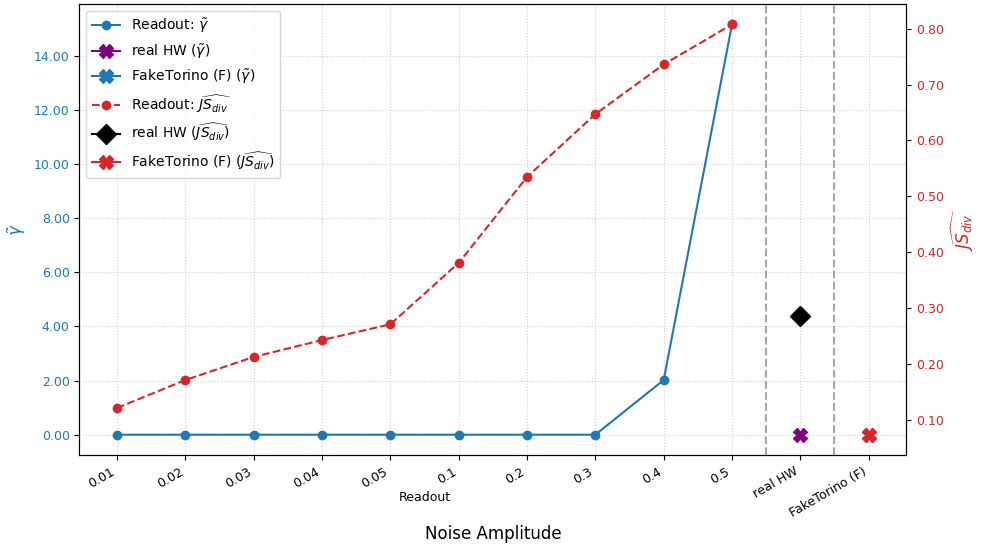}
\caption{{\bf Experiment~2 -- Divergence.} Plot of \(\widehat{JS}_{\mathrm{div}}\) and the flip-rate \(\tilde{\gamma}\) under readout noise.}
\label{fig:exp2_2Metrics}
\end{figure}

\subsubsection{Observed behavior}

Figures~\ref{fig:exp2_winner_agreement} and~\ref{fig:exp2_2Metrics} show a smooth, monotonic increase of $\widehat{JS}_{\mathrm{div}}$ with readout-noise amplitude, from a low-divergence regime at small $p$ to a substantially deformed societal distribution at the largest noise levels. Throughout the range $p\le 0.3$, the normalized flip rate remains identically zero and winner–agreement stays perfect ($\Gamma_{\mathrm{win}}=1$), indicating that the Condorcet winner is completely stable even as the distribution gradually drifts away from the noiseless reference. At $p=0.4$ the flip rate becomes modestly non-zero and the agreement remains essentially perfect, while at the extreme amplitude $p=0.5$ both the flip rate and $\widehat{JS}_{\mathrm{div}}$ become large and $\Gamma_{\mathrm{win}}$ collapses, signaling a qualitative change in the majority structure.

The real-hardware point exhibits $\Gamma_{\mathrm{win}}=1$, $\tilde{\gamma}=0$, and a moderate divergence $\widehat{JS}_{\mathrm{div}}$ that lies well within the trend traced by the readout sweep. The FakeTorino backend is even closer to the noiseless baseline, effectively matching the low-noise end of the readout curve. Both device-level points sit on the same stable trend as the simulator sweep and are best interpreted as additional noisy operating points, rather than as evidence of any intrinsic instability in the QMR constitution itself.

\noindent\textbf{Summary}

This decisive Condorcet profile with wide pairwise majorities is highly
noise-robust.
Across all practically relevant readout and device-noise levels ($p\le 0.3$) we
observe perfect winner–agreement $\Gamma_{\mathrm{win}}=1$ and a vanishing
normalized flip rate $\tilde{\gamma}=0$, while
$\widehat{JS}_{\mathrm{div}}$ grows in a smooth and controlled fashion from a
small value at $p=0.01$ to a moderate deformation at intermediate noise.
At $p=0.4$ the divergence continues to increase and the flip rate becomes
modestly non-zero, yet $\Gamma_{\mathrm{win}}$ remains essentially perfect.
Only at the intentionally extreme setting $p=0.5$ does the system exhibit both
a large flip rate and a near-complete loss of agreement with the classical
Condorcet winner, signalling a qualitative change in the majority structure.

Real hardware and FakeBackend runs align qualitatively with the simulator
trends, differing mainly in their absolute divergence values: both device-level
points sit on the same stable branch as the low-to-moderate readout amplitudes
and retain $\Gamma_{\mathrm{win}}=1$ and $\tilde{\gamma}=0$.
Taken together, these observations confirm that for strong classical
majorities the QMR constitution preserves both the winner and the underlying
SCC structure under realistic noise, and only breaks down under deliberately
large readout perturbations.

\subsection{QMR2-inspired protocol}
\label{Results_QMR2}

In contrast to the previous subsections, which evaluate the full QMR
constitution using three complementary stability metrics
($\Gamma_{\mathrm{win}}$, $\tilde{\gamma}$, and $\widehat{JS}_{\mathrm{div}}$)
across classical, simulator, FakeBackend, and hardware runs, this
subsection is deliberately more exploratory and schematic. Here we
depart from the full QMR constitution—comprising the Tarjan-based SCC
construction, the GMS modification, and the
EU steps—and instead study a simplified,
QMR2-inspired setting whose purpose is to isolate the effect of
multi-voter entanglement on simple majority outcomes and draw rates,
both ideally and under a basic local-noise model. The
aggregation rule is a mini-round majority vote, and performance is
reported via frequency histograms of winners and draws rather
than metrics used for QMR.

Operationally, we instantiate this QMR2-inspired setting as a quantum constitution designed to test whether groups of voters can
synchronize their preferences to produce an “optimal” classical outcome.
We emphasize that the cases studied here are not the ones that appear in BYH, 
rather, addressing the test case scenario comprising many groups of voters, with potentially synchronized (separable or entangled), or random ballots. 
We consider a voting system
with \(m\) candidates (so \(m!\) possible strict rankings) and a group of
\(n\) voters participating in each round. In each round, these
\(n\) voters submit their preferences to a quantum device in the form of
a joint state, which may be either separable or entangled, thereby
allowing up to \(n\) voters to become coherently correlated in that
round. Each iteration implements a ``mini'' majority rule. For an arbitrary outcome represented by a classical string
\[
S = \{0,1,2,1,\ldots,1\}.
\]
The winner of that mini round is the preference that appears most frequently among the $n$ voters participating in that iteration. We then count the frequency of these mini-round winners across all iterations to determine the final group winner. This constitution, therefore, functions as a quantum analogue of a repetition code: it aggregates several noisy or uncertain inputs to infer a stable collective outcome.

We compare three scenarios: (i) voters sharing an entangled
GHZ-type preference state as defined in
Eq.~(\ref{Eq:GHZ}), (ii) voters prepared in separable superpositions as defined in Eq.~(\ref{eq:sep}), and (iii) fully random voters, where each voter
independently chooses a preference uniformly at random. If, in a given iteration, the mini majority rule fails to identify a unique winner, the outcome is recorded as a draw and denoted by $-1$.

\begin{figure}[ht]
    \centering
    \begin{subfigure}[t]{0.48\linewidth}
        \centering
        \includegraphics[width=\linewidth]{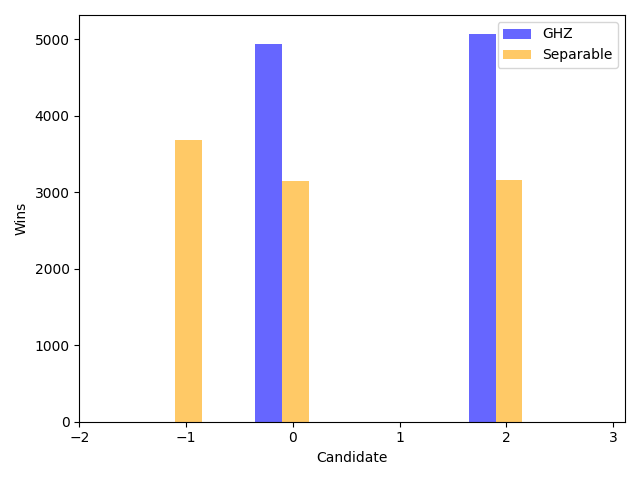}
        \caption{{\bf Separable and GHZ scenarios without random voters.} Draw events occur only in the separable scenario.}
        \label{fig:sepvsghzm4n3}
    \end{subfigure}
    \hfill
    \begin{subfigure}[t]{0.48\linewidth}
        \centering
        \includegraphics[width=\linewidth]{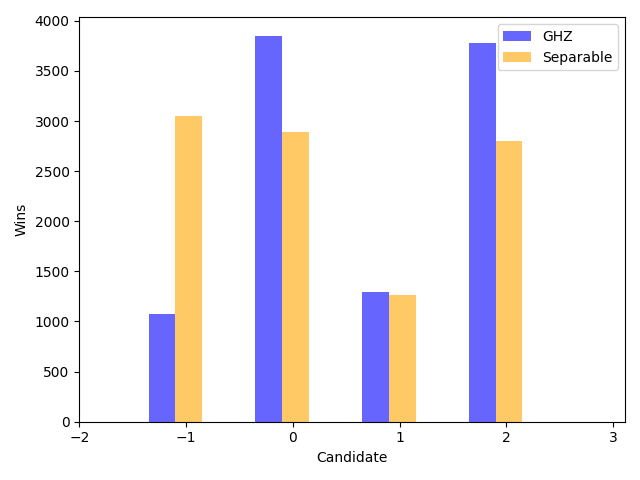}
        \caption{{\bf Separable and GHZ scenarios with random voters}. 1:1 ratio of random and matched voters in the same iteration.}
        \label{fig:sepvsghzm4n3_random}
    \end{subfigure}
    \caption{QMR2-inspired toy constitution with three candidates \(C = \{0,1,2\}\) and mini-rounds of \(n\) voters. For each configuration of \(n\) voters, a single outcome is obtained and the process is repeated \(10{,}000\) times. We compare separable states defined in Eq.~(\ref{eq:sep}) and the GHZ state defined in Eq.~(\ref{Eq:GHZ}) under identical settings, which shows that, on average, they yield the same expectation value with respect to the preference operator defined in Eq.~(\ref{eq:Psi_gamma_ket}). However, when the voters are entangled, the draw outcome is trivially absent.}
    \label{fig:QMR_combined}
\end{figure}

For the simulations presented here, we use $m = 3$ candidates, $k = 4$ voters per iteration, and $10000$ iterations (corresponding to $40,000$ voter instances in total). The results are shown in Fig.~\ref{fig:QMR_combined}.  The main difference between the separable voters and the entangled voters is that for the entangled voters, draw outcomes are completely absent. This follows from the fact that each group of $n$ voters behaves coherently, producing unanimous outcomes by construction.

\subsubsection{Noisy QMR2 system}
Now we add local noise on each encoded qubit (local bit flip). For example, if the state $\ket{L}$ is encoded, we know that there is a representation in the preference basis that is isomorphic to the qubits representation, which is affected by the noisy channel with some probability $p$ that is fixed in all the local quantum channels.
The quantum channel is defined as:
\begin{equation}
    \mathcal{M}(\rho) = \sum_i F_i\rho F^{\dagger}_i = (1-p)\rho + p\sigma_x \rho \sigma_x
\end{equation}

\begin{figure}[H]
    \centering
    \includegraphics[width=0.8\linewidth]{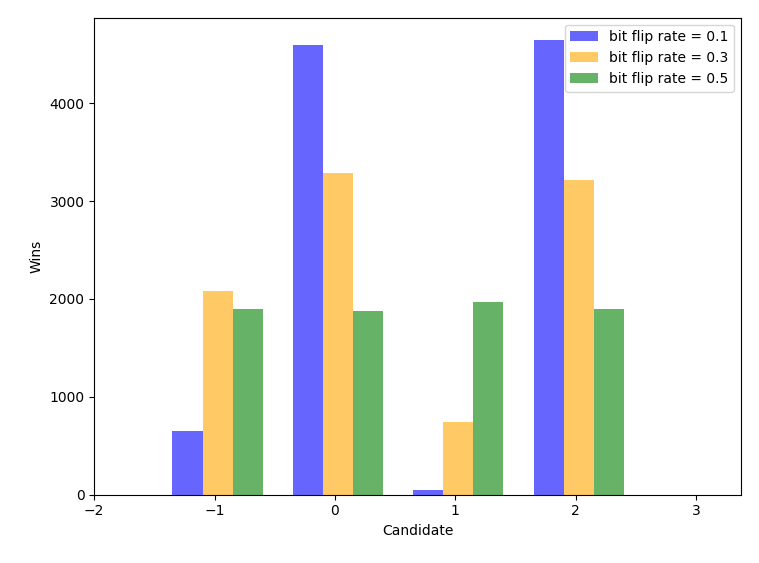}
    \caption{{\bf Noisy QMR2-inspired constitution.} Results for local bit-flip probabilities \(p=0.1\) (blue), \(p=0.3\) (yellow), and \(p=0.5\) (green). Increasing the noise level drives the GHZ state toward behavior which is indistinguishable from sampling a pool of random voters, leading to an inconclusive winner distribution at high \(p\).}
    \label{fig:qmr2_noisy}
\end{figure}

Where $p$ is the probability for the Pauli X operator. More explicitly the Kraus operators are defined as: $F_1 = \sqrt{1-p} \mathds{1}, F_2 = \sqrt{p}\sigma_x$.
We test three noise levels, \(p \in \{0.1,0.3,0.5\}\), as shown in Fig.~\ref{fig:qmr2_noisy}. Increasing bit-flip noise progressively destroys the coherent GHZ correlations, driving the entangled voter block toward behavior indistinguishable from a randomly mixed voter population.

The results show that the GHZ state loses coherence as the noise magnitude increases, exactly as expected from standard quantum information theory. In particular, when $p = \frac{1}{2}$, the state approaches the maximally mixed limit. Furthermore, the behaviour of the random voters introduced in the previous section is consistent with the effect of sampling from a uniform random voter pool.

An advantage of the GHZ state is that, after post-selecting the events without draws (which never occur in the entangled case), the relative proportions of the outcomes $L$ and $\bar{L}$ are largely preserved. This stability reflects the strong correlation structure inherent in GHZ-type states, even in the presence of moderate noise.

From the perspective of quantum social choice, these toy experiments complement the QMR hardware runs: they illustrate that entanglement can enforce highly coordinated, draw-free outcomes in ideal conditions, but that this benefit is quickly eroded once local noise reaches moderate levels, in contrast to the more profile-dependent yet quantitatively robust behavior observed for QMR.


\section{Discussion, Conclusion and Future Work}
\label{section_conclusion_future_work}

\subsection{Summary of main findings}

Our results can be grouped into two main messages. First, for the QMR constitution implemented on five voters and three candidates, we observe a robust but profile–dependent response to readout and device noise. For both representative profiles (Experiments~1 and~2), the readout-noise sweeps show a smooth, monotonic increase of the divergence $\widehat{JS}_{\mathrm{div}}$ as the noise amplitude $p$ grows from $0.01$ to $0.5$, while the normalized Condorcet–winner flip rate $\tilde{\gamma}$ remains exactly zero and the winner–agreement $\Gamma_{\mathrm{win}}$ stays at $1$ throughout the entire range $p\le 0.4$. In this regime, QMR preserves the classical Condorcet winner and the SCC structure of the majority digraph, while noise primarily manifests as a broadening of the societal ranking distribution. The two profiles differ only in their behaviour at the most extreme point $p=0.5$: in Experiment~2, the decisive Condorcet profile starts to show moderate winner flips and a larger $\widehat{JS}_{\mathrm{div}}$, whereas in Experiment~1, which has narrower pairwise margins and a more fragile SCC structure, the system undergoes a sharper transition to a different top set with a larger top SCC and a higher flip rate. Even in this stress–test regime, however, the societal outcome does not become random; it converges to a reproducible noisy equilibrium. Taken together, these two qualitatively different profiles demonstrate that QMR remains stable over a wide range of realistic readout and device noise levels, with profile–dependent deviations appearing only under deliberately extreme noise. The real–hardware and FakeBackend data points lie on the same stable trend as the simulator sweeps, with $\Gamma_{\mathrm{win}}=1$ and $\tilde{\gamma}=0$, indicating that realistic device–level noise places QMR well inside the robust regime in which the classical Condorcet winner is retained and the internal SCC topology is preserved.

Second, our QMR2-inspired entanglement experiments show that entangled voter blocks can significantly modify outcome statistics in ideal conditions. In the noiseless limit, GHZ-type groups of voters essentially eliminate draw outcomes and bias the majority toward a preferred alternative, even in the presence of additional separable or random voters. Under moderate local bit-flip noise, however, this advantage is fragile: the winner frequencies and draw rates revert to those of separable or random-voter baselines, and GHZ and separable populations become increasingly indistinguishable at the level of societal statistics.

\subsection{Limitations and scope}

Our study has several limitations that naturally bound the scope of the conclusions.

\begin{itemize}
    \item \textbf{System size.} We focus on $n=5$ voters and $m=3$ candidates, so that $m!=6$ rankings per voter and majority digraphs remain tractable, and SCC enumeration is straightforward. Larger electorates or more candidates would increase both the classical analytic complexity (e.g., in SCC-based mixing and linear extension counting) and the qubit requirements for a direct preference encoding.
    \item \textbf{Noise models} The circuits used in this work are simple initialize–measure circuits with a limited entangling structure. Our explicit noise studies emphasize readout noise; we do not model crosstalk, leakage, non-Markovian effects, or long-depth error accumulation. The observation that readout dominates is therefore specific to this regime.
    \item \textbf{QMR2-inspired nature of the entanglement experiments.} The QMR2-inspired protocol is a toy variant: it borrows the strategic spirit of the BYH QMR2 scenarios (entangled blocs, interference between tactics) but does not implement the full BYH QMR2 constitution. Consequently, we do not claim Arrow-type properties for this protocol and treat it purely as a controlled setting for probing multi-voter quantum correlations under noise.
\end{itemize}

These constraints are natural for a first hardware-based robustness study of QMR, but they also mean that our conclusions should be interpreted as evidence about the behavior of small, readout-dominated NISQ implementations rather than as general asymptotic statements about quantum social choice in large electorates.

\subsection{Implementation considerations: circuit layout and encoding choices}
\label{subsec:discussion_circuit}

The QMR experiments reported in Sec.~\ref{section_results} use the
15-qubit, per-voter layout described in Sec.~\ref{section_methods}.
Here we justify this choice and contrast it with a compressed 3-qubit
reuse scheme.

A central question in our study is why implement a quantum circuit at all when one could, in principle, compute the induced societal state analytically as a distribution or density operator~$\rho_{\mathrm{soc}}$ and then derive all observables (Condorcet winner, majority digraph, SCC structure, ranking distributions) without using any quantum simulator or hardware. The analytical route is attractive: it is exact, inexpensive, and fully reproducible.

However, it hides precisely the effects we aim to measure:  
(i)~\textbf{Readout error}, which we probe directly by sweeping a controllable readout–noise $\eta_{\mathrm{ro}}$ on the simulator and observe again on hardware; and  
(ii)~\textbf{Device implementation effects} arising from mapping the same circuit onto an actual connectivity graph and native gate set.  
We study~(ii) by running the identical compiled circuit on an IBM FakeBackend and on real hardware. We compare these runs to the noiseless-simulator baseline and summarize the differences using $\widehat{JS}_{\mathrm{div}}$, $\tilde{\gamma}$, and $\Gamma_{\mathrm{win}}$ computed over repeated batches. In this way, the observed gaps reflect readout and routing–related implementation noise rather than differences in the analytical model.

Our metrics—winner agreement, flip behavior, and distributional divergence—are ultimately statements about robustness under implementation. Those properties live in the end-to-end quantum pipeline; a purely analytic computation of $\rho_{\mathrm{soc}}$ without explicit circuit instantiation would therefore overestimate robustness. The circuit instantiation is thus not merely a sampling platform but an empirical interface that ties the theoretical constitution (QMR/QMR2-inspired) to the physical substrate on which it may ultimately operate.

\medskip
For $n=5$ voters and $m=3$ candidates ($m!=6$ rankings per voter), we compared two encoding strategies: a per-voter layout with three qubits per voter (total~15~qubits) and a 3-qubit reuse scheme that applies a single 3-qubit register sequentially across the five voters using mid-circuit measurement and reset.  
The 15-qubit layout offers clear voter locality---each register corresponds to one voter’s ranking---so superpositions are easy to interpret and decoding is a straightforward per-register lookup.
It can also natively support QMR2-inspired experiments with entanglement across voters.
Its main cost is a higher qubit count: measuring more qubits in parallel amplifies the impact of per-qubit readout errors, so the probability that an entire 15-bit outcome is error-free is lower than for a 3-qubit register. In practice, this can increase the effective readout error on profiles and may require more shots to reach the same statistical precision. Extra two-qubit gates or added depth are not inherent and appear only when multi-qubit interactions are intentionally introduced or routed.

The 3-qubit reuse scheme uses fewer qubits and fits smaller devices, but it loses the ability to represent the full joint profile simultaneously. In practice, it measures and resets after each voter, so no coherent joint profile amplitudes are preserved. As a result, it does not perform a projective measurement of the global societal state $\rho_{\mathrm{soc}}\in\mathbb{C}^{6^5\times 6^5}$; instead, it collects per-voter outcomes and stitches them together classically. This alters the noise characteristics and extends execution time over multiple measure–reset rounds (increasing latency and opportunities for error accumulation), even if each round remains shallow. Decoding back to per-voter rankings is also less transparent and more error-prone.

\medskip
Balancing these considerations, we conclude that the 15-qubit, per-voter layout is the appropriate choice for experiments that (i)~must support QMR2’s engineered entanglement across voters, (ii)~require transparent, auditably correct decoding into majority graphs and SCCs, and (iii)~aim to attribute observed instabilities to genuine physical and architectural causes rather than to encoding artifacts.  
While a 3-qubit compressed design may suffice for limited, non-entangled, noiseless demonstrations of QMR’s analytic predictions via~$\rho_{\mathrm{soc}}$, it undermines the very hypotheses we test when moving to noise sweeps and hardware.  
In short, the circuit is necessary to expose implementation-level sensitivities, and the 15-qubit voter-factorized encoding is necessary to reveal them in a way that preserves the logical structure of the constitution, the interpretability of measurements, and the validity of our robustness metrics.

More broadly, these considerations apply to any protocol that requires
transparent reconstruction of joint preference profiles and controlled
use of inter-voter entanglement, suggesting that qubit-efficient reuse
schemes must be adopted with caution when the collective state itself is
the object of study.

\subsection{Future work: quantum error correction and noisy Arrow analysis}

A natural next step is to move from noise-aware simulations toward explicit quantum error correction for voting circuits executed on real hardware. In the present work we characterize how device-level noise and readout imperfections distort QMR and QMR2 outcomes, but we do not yet attempt to protect the voting process using established fault-tolerant techniques. Future experiments could encode the voting registers into standard codes such as repetition, surface, or small-distance stabilizer codes, and implement syndrome extraction and recovery directly within the QMR circuit. This would enable a systematic comparison between unencoded and encoded quantum voting, quantifying not only the improvement in Condorcet-winner stability and majority-digraph statistics, but also the overhead in qubit count, circuit depth, and sampling time required to achieve a given level of social-choice robustness.

Building on such hardware-level demonstrations, one can then treat the quantum voting pipeline as a concrete testbed for standard fault-tolerance concepts. For example, one may study how logical error rates after decoding translate into effective social-choice noise models such as random ballot replacement, voter flips, or structured abstention patterns. This mapping would allow a direct calibration between physical error channels (depolarizing, dephasing, readout bias, and qubit loss) and their impact on majority cycles, top-cycle stability, and winner-agreement metrics. A particularly interesting direction is to investigate whether moderate-distance codes on current superconducting or trapped-ion platforms can already push the system into a regime where the logical social outcome is significantly more stable than the raw physical measurement, even for relatively small electorates.

In parallel with these experimental directions, there is substantial room for analytical work on Arrow-style impossibility properties in noisy and lossy regimes. Classical Arrow theorems are formulated for ideal, noise-free preference profiles, whereas real hardware inevitably introduces errors, erasures, and incomplete information. Future theoretical work could formalize a noisy version of Arrow’s framework in which voters’ rankings are perturbed by stochastic channels or partially deleted, and then prove under what conditions the key properties (such as weak Pareto efficiency, non-dictatorship, and independence of irrelevant alternatives) are preserved, weakened, or effectively regularized by noise. For QMR, one could seek rigorous bounds that relate the strength and structure of physical noise to the probability of observing Arrow-type paradoxes, or to the persistence of Condorcet cycles in the large-voter limit.

Finally, combining these two strands suggests a broader research program in which rigorously characterized error models and error-corrected circuits are studied within a unified social-choice framework. On the one hand, experiments on encoded voting circuits would supply empirical data about how real noise processes and recovery procedures distort or stabilize collective rankings. On the other hand, analytical results on Arrow properties under noise and loss would provide principled performance targets and no-go theorems for any putative “fair” quantum voting architecture. Together, these developments would elevate quantum voting from a proof-of-principle algorithm to a setting where questions of fault tolerance, impossibility, and democratic robustness can be confronted simultaneously and with mathematical precision.\\

{\bf Data and Code Availability.} The datasets generated and analyzed during the current study, including raw outcome logs (Condorcet-winner labels, majority-digraph strongly connected components, and per-batch values of the Jensen--Shannon divergence $\widehat{JS}_{\mathrm{div}}$), are available from the corresponding author upon reasonable request. The quantum-voting simulation and data-analysis code used to implement the QMR and QMR2-inspired protocols, compute the summary metrics $\Gamma_{\mathrm{win}}$ and $\tilde{\gamma}$, and generate the figures in this work is likewise available from the corresponding author upon reasonable request.\\

{\bf Acknowledgments.} This work was supported by the European Union’s Horizon Europe research and innovation programme under grant agreement No. 101178170 and by the Israel Science Foundation under grant agreement No. 2208/24.
\appendix
\section*{Appendix}

The appendices present technical definitions and derivations that support the main text. 
Appendix~\ref{app_classical_choice_theory__notations} and Appendix~\ref{app_Arrow's_properties} 
summarize the classical ranked-voting framework and Arrow's theorem. 
Appendix~\ref{app_BYH_quantum_Arrow's_properties} and Appendix~\ref{app_BYH_quantum_QMR} discuss 
the quantum generalizations and a structured description of the QMR and QMR2 constitutions. 
Finally, we include additional figures and exploratory results that complement, but are not essential 
to, the main narrative.
\section{Classical social choice theory: ranked-voting and Arrow's theorem}
This appendix provides a compact reference for the classical ranked-voting framework used 
throughout the paper. We first collect the basic notations and order-theoretic concepts 
(Table~\ref{tab:classical_notation}), then restate Arrow's axioms and give a streamlined 
proof of the impossibility theorem that underpins our later quantum generalizations.
\subsection{Notations}\label{app_classical_choice_theory__notations}
\begin{table}[h!]
    \centering
\begin{tabular}{|m{0.18\textwidth}|m{0.55\textwidth}|m{0.22\textwidth}|}
\hline
    Notion& Definition& Comments\\ 
    \hline
    \hline
    \multicolumn{3}{|c|}{Basic notations}\\
    \hline
    Cartesian product &
    $ A\times A=\{(x,y):x,y\in A\}$ & Set of all possible pairs\\
    \hline
    Relation &
    $R\subseteq A\times A$ & \\
    \hline
    Universal relation &
    $U_A= A\times A$ & \\
    \hline
    \multicolumn{3}{|c|}{Families of relations satisfying certain conditions (orders)}\\
    \hline
    Complete & 
   
    $\text{Comp}(A)= \{
    R\subseteq U_A:\forall x\ne y\in A, (x,y)\in R\lor(y,x)\in R
    \}$
    &
    All complete orders\\
    \hline
    Reflexive & 
    
    $\text{Refl}(A)=\{
    R\subseteq U_A:\forall x\in A, (x,x)\in R\}$
    &
    All reflexive orders\\
    \hline
    Irreflexive& 
    $\text{Ir}(A)=\{R\subseteq U_A:\forall x\in A, (x,x)\notin R\}$
    &
    All irreflexive orders\\
    \hline
    Transitive & 
      $\text{Trans}(A)=\newline 
    \{R\subseteq U_A:\forall x,y,z\in A,
    (x,y)\in R\land (y,z)\in R
    \Rightarrow
    (x,z)\in R\}$
    &
    All transitive orders\\
    \hline
    Symmetric& 
    $\text{Sym}=\{R\subseteq U_A:\forall x, y\in A, 
    (x,y)\in R\Rightarrow(y,x)\in R
    \}$
    &
    All symmetric orders\\
    \hline
    Antisymmetric & 
    $\text{AntiSym}=
    \newline 
    \{R\subseteq U_A:\forall x,y\in A,
    (x,y)\in R \land (y,x)\in R\Rightarrow x=y\}
    \newline \text{or}\newline 
    \{R\subseteq U_A:\forall x\ne y\in A,
    (x,y)\in R \Rightarrow (y,x)\notin R\}
    $
    &
    All antisymmetric orders
    \\
    \hline
    Asymmetric & 
    $\text{ASym}=\{R\subseteq U_A:
    \forall x,y \in A,(x,y)\in R\Rightarrow(y,x)\notin R
    \}$
    &
    All asymmetric orders
    \\
    \hline
    Strict partial&  
    $l(A)\in
    \text{Trans}(A)\cap\text{Asym}(A)$
    & Transitive and asymmetric \\
    \hline
    Strict linear  &
    $\mathcal{L}(A)=
    \text{Comp}(A)\cap\text{Trans}(A)\cap\text{Asym}(A)$
    & Complete, transitive and asymmetric\\
    \hline
    Weak &
    $\textbf{R}(A)=
    \text{Comp}(A)\cap\text{Trans}(A)\cap\text{Refl}(A)$
    &Complete, transitive and reflexive\\
    \hline
    \hline   
    \multicolumn{3}{|c|}{Voting schemes}\\
    \hline
    Set of alternatives & $A=\{a_1,\hdots,a_m\}$ & $m$ alternatives\\
    \hline
    Set of voters & $V=\{v_1,\hdots,v_n\}$ & $n$ voters\\
    \hline
    Coalition & $C\subseteq V$ & Any subset of the voters set.\\
    \hline
    Strict preference & $(x,y)=x\succ y$ & $x$  is strictly preferred to  $y$\\ 
    \hline  
    Non-strict preference & $(x,y)=x\succsim y$ & $x$ is \textit{at least as good as} $y$ \\ 
    \hline
    Indifference& $x\ne y,x\succsim y\land y\succsim x\Rightarrow x\sim y$& Tie between $x$ and $y$\\
    \hline
    Voter $i$'s preference & $L_i=\left(a_{i(max)}\succ\cdots\succ a_{i(min)}\right)\in\mathcal{L}(A), i\in V$ & Linear strict order\\
    \hline
    Profile & $\vec{L}=(L_1,\hdots,L_n)\in\mathcal{L}(A)^n$ & All voters' preferences\\
    \hline
    Social Welfare Function (SWF) & $f(P)=R_\text{soc}(A):\mathcal{L}(A)^n\to \textbf{R}(A)$ & 
    The social preference is a weak, non-strict order when voters' preferences are linear and strict\\
    \hline
\end{tabular}
  \caption{Classical ranked-voting notations used in the paper.}
    \label{tab:classical_notation}
\end{table}
\newpage
\subsection{Arrow's properties}\label{app_Arrow's_properties}
We would like to briefly introduce, using the classical choice theory notations, appearing in Appendix~\eqref{app_classical_choice_theory__notations}, the ranked-voting scheme's properties according to Arrow~\cite{Arrow1951}:
\begin{enumerate}
    \item Transitivity: Society's preferences $a\succsim_\text{soc} b$ and $b\succsim_\text{soc} c$ imply $a\succsim_\text{soc} c$.  
    \item Unanimity: 
    If every voter prefers $a$ to $b$, then society's preference should be the same:\\ 
    $\forall i\in V, a\succsim_i b $ implies $ a\succsim_\text{soc}b$  for any $a,b\in A$.     
    
    \item Independence of irrelevant alternatives (IIA): The relative ranking of any two candidates, say $a$ and $b$, should not be influenced by the ranking of any other candidate. Given two preference profiles (different elections), $\vec{L}=(L_1,...,L_n)$ and $\vec{L}'=(L'_1,...,L'_n)$, if all the individual preferences regarding any two alternatives coincide in the profiles $\vec{L}$ and $\vec{L}'$, then the social preferences coincide as well.\\ 
    $\forall i\in V, r_i^{ab}={r'}_i^{ab}\Rightarrow r_\text{soc}^{ab}={r'}_\text{soc}^{ab}$, where $r_i\in\{\succ,\prec\}$ and $r_\text{soc}\in\{\succsim,\precsim\}$.
    \item Dictatorship: Society prefers $a$ to $b$ iff some voter $i$ prefers $a$ to $b$, 
    $\exists i: a\succ_i b  \hspace{1mm}\Leftrightarrow\hspace{1mm} a\succsim_\text{soc}b$.
\end{enumerate}

\section{Quantum ranked-voting}
This appendix translates the classical ranked-voting framework into a quantum setting. 
We define the preference Hilbert space, the space of density operators, and the notion of a 
quantum social welfare function (QSWF), together with the measurement statistics that play 
the role of ranking probabilities in the main text.

\subsection{Notations}
\begin{table}[h]
    \centering
\begin{tabular}{|m{0.25\textwidth}|m{0.40\textwidth}|m{0.3\textwidth}|}
    \hline
    \multicolumn{3}{|c|}{Basic relevant notations in quantum mechanics} \\
    \hline
    Notion& Definition& Comments\\ 
    \hline
    Pure state& $\ket{\psi}=\sum_i c_i\ket{u_i}\in \mathcal{H}$&\\
    \hline
     Mixed state -- density operator &$\rho=\sum_k p_k\ket{\psi_k}\bra{\psi_k}\in\mathcal{B}(\mathcal{H})$
     \newline
     $\hspace{2mm}\rho^\dagger=\rho,\hspace{2mm}\rho\succeq 0,\hspace{2mm}\text{Tr}\rho=1$&Here $\mathcal{B}$ is a Banach space\\
    \hline
    Projector onto a subspace $h\subseteq\mathcal{H}$&
    $\Pi^h=\sum_{\ket{\phi}\in h}\ket{\phi}\bra{\phi}, \hspace{5mm}
    h=\Pi^h\mathcal{H}\Pi^h$&\\
    \hline
    Support subspace of a density operator &$\text{supp}(\rho)=\text{span}\{\ket{\phi}\in \mathcal{H}:\rho\ket{\phi}\ne0\}\subseteq\mathcal{H}$&\\
    \hline
    Support of $\rho$ on $h\subseteq\mathcal{H}$&$p(h)=\text{Tr}(\Pi^h\rho)$&The probability of the measurement outcome of $\rho$ to belong to $h$\\
    \hline
    $\rho$ has support only on $h$&$\rho=\Pi^h\rho\Pi^h$&\\
    \hline
\end{tabular}
  \caption{Basic quantum-mechanical notation used in the quantum voting framework.}
    \label{tab:qm_notation}
\end{table}

\begin{table}[h]
    \centering
\begin{tabular}{|m{0.2\textwidth}|m{0.45\textwidth}|m{0.3\textwidth}|}
    \hline
    \multicolumn{3}{|c|}{Quantum voting} \\
    \hline
    Notion& Definition& Comments\\ 
    \hline
    Preference space& 
    $\mathcal{H}_A:=\text{span}(B_L),\hspace{5mm} B_L=
    \{\ket{L}:L\in\mathcal{L}(A)\}$
    & 
    $L$'s are linear orders on $A$, and $B$ is a preference basis\\
    \hline
    Classical profile&
    $\vec{L}=
    (L_1,\ldots,L_n)\in\mathcal{L}(A)^n$
    &Same as in Table~\ref{tab:classical_notation}\\
    \hline
    Set of density operators on $\mathcal{H}_A$&$\mathcal{D}(\mathcal{H}_A)=\{\rho\in\mathcal{B}(\mathcal{H}_A)|\rho^\dagger=\rho, \rho\succeq 0, \text{Tr}(\rho)=1\}$& Analog of $\mathcal{L}(A)$\\
    \hline
    Multi-voter joint state &$\sigma_\text{soc}\in\mathcal{D}(\mathcal{H}_A^{\otimes n})$&Analog of a profile $P\in\mathcal{L}(A)^n$\\
    \hline
    Individual voter's state&$\rho_i=\text{Tr}_{V\setminus i}\,\,\sigma_\text{soc} \in\mathcal{D}(\mathcal{H}_A)$&Reduced individual density matrix\\
    \hline
    Quantum profile&$\mathcal{P}=\{\rho_i,\ldots,\rho_n\}\in\mathcal{D}(\mathcal{H}_A^{\otimes n})$&\\
    \hline
    Product state (uncorrelated voters)& $\sigma_\text{soc}=\rho_1\otimes...\otimes\rho_n$ &In this case $\rho_i=\sum_k p_k\ket{\psi_k}\bra{\psi_k}$\\
    \hline
    Subspace of partial linear order $l(A)$ &
    $\mathcal{G}^{l (A)}=\text{span}\{\ket{L}\in B_L: l(A)\}
    $
    &\\
    \hline
    Pair preference subspace &
    $\mathcal{G}^{a \succ b}=\text{span}\{\ket{L}\in B_L: a \succ b)\}$&
    \\
    \hline
    Preference basis element& $\mathcal{G}^L=\ket{L}$&\\
    \hline
    Partial order projector&
    $\Pi^{\mathcal{G}^{l(A)}}=\sum_{\ket{L}\in \mathcal{G}^{l(A)}}\ket{L}\bra{L},
    \hspace{3mm}
    \newline
    \mathcal{G}^{l(A)}=    \Pi^{\mathcal{G}^{l(A)}}\mathcal{H}_A\Pi^{\mathcal{G}^{l(A)}}$&\\
    \hline
    Quantum SWF&
    $\mathcal{E}(\sigma_\text{soc})=\rho_\text{soc},
    \hspace{5mm}
    \mathcal{E}:\mathcal{D}\left(\mathcal{H}_A^{\otimes n}\right)\rightarrow\mathcal{D}(\mathcal{H}_A)$&
    Convex-linear CPTP map \\
    \hline
    Quantum SWF measurement&$p_\text{soc}(L)=\text{Tr}(\ket{L}\bra{L}\rho_\text{soc})$&Probability of the linear order $L$ outcome\\
    \hline
\end{tabular}
  \caption{Notation for quantum ranked-voting and quantum social welfare functions.}
    \label{tab:qvoting_notation}
\end{table}
\newpage
\subsection{Quantum version of Arrow's properties}\label{app_BYH_quantum_Arrow's_properties}
In this section we collect the quantum analogues of Arrow's classical axioms, following 
the formulation in Ref.~\cite{BaoHalpern2017}. The goal is not to introduce new conditions, but to provide 
a precise language for discussing when a quantum social welfare function satisfies quantum 
transitivity, quantum unanimity, quantum IIA (QIIA), and quantum dictatorship.

We follow here the definitions given in a paper by Sun et al.~\cite{Sun2021}. In the following we use the probability of the preference $l(A)$ for voter $i$/society, in profile $\mathcal{P}$, given by $p^\mathcal{P}_{i/ \text{soc}}(l(A))=\text{Tr}_{\ne i/\text{soc}}(\Pi^{l(A)}\rho^\mathcal{P}_{i/ \text{soc}})$.

\begin{enumerate}
    \item Quantum Transitivity:\\ The measurement of $\rho_\text{soc}$ in the preference basis, $B=\{\ket{L}:L\in\mathcal{L}(A)\}$, yields some $\ket{L}$, with probability $p_\text{soc}(L)$, which is transitive, by definition.
    \item Quantum Unanimity: 
    \begin{enumerate}
        \item Sharp: $\forall i\in V,\forall a,b\in A \hspace{3mm}
        p_i(a\succ b)=1\Rightarrow
        p_\text{soc}(a\succ b)=1$
        \item Unsharp: $\forall i\in V,\forall a,b\in A\hspace{3mm} 
        p_i(a\succ b)>0\Rightarrow
        p_\text{soc}(a\succ b)>0$
        \end{enumerate}
    A quantum SWF $\mathcal{E}$ satisfies the quantum unanimity condition if it satisfies both sharp and unsharp conditions.
    \item Quantum IIA (QIIA):\\
    Given any two quantum preference profiles -- multi-voter joint states, $\sigma_\text{soc}$, and $\sigma'_\text{soc}$, and the resulting society's preferences $\rho_\text{soc}=\mathcal{E}(\sigma_\text{soc})$, and $\rho'_\text{soc}=\mathcal{E}(\sigma'_\text{soc})$, 
    \begin{enumerate}
        \item Sharp: If $\forall i\in V, \forall a,b\in A,\hspace{3mm}\\
        p_i(a\succ b)=p'_i(a\succ b)$
         and 
        $p_i(b\succ a)=p'_i(b\succ a)$, then $p_\text{soc}(a\succ b)=1$ implies  $p'_\text{soc}(a\succ b)=1$
        \item Unsharp: If $\forall i\in V,\forall a,b\in A,$\\
        $\mathcal{P}_i(a\succ b)=\mathcal{P}'_i(a\succ b)$
        and
        $p_i(b\succ a)=p'_i(b\succ a)$,
        then $p_\text{soc}(a\succ b)>0$ implies  $p'_\text{soc}(a\succ b)>0$
    \end{enumerate}
    A quantum SWF $\mathcal{E}$ satisfies the QIIA condition if it satisfies both sharp and unsharp conditions.
    
    \item Quantum Dictatorship:
    \begin{enumerate}
        \item Sharp: 
        $\exists i,\forall \mathcal{P}\in\mathcal{D}(\mathcal{H}_A^{\otimes n}),\forall a,b\in A, \hspace{3mm}
        p^\mathcal{P}_i(a\succ b)=1
        \hspace{3mm}\Leftrightarrow \hspace{3mm}
        p^\mathcal{P}_\text{soc}(a\succ b)=1$
        \item Unsharp:
        $\exists i,\forall \mathcal{P}\in\mathcal{D}(\mathcal{H}_A^{\otimes n}),\forall a,b\in A, \hspace{3mm}
        p^\mathcal{P}_i(a\succ b)>0
        \hspace{3mm}\Leftrightarrow \hspace{3mm}
        p^\mathcal{P}_\text{soc}(a\succ b)>0$
    \end{enumerate}
    A quantum SWF $\mathcal{E}$ satisfies the quantum dictatorship condition if it satisfies both sharp and unsharp conditions.
\end{enumerate}

\subsection{BYH quantum SWFs}
\subsubsection{QMR}\label{app_BYH_quantum_QMR}
The QMR constitution can be expressed as a sequence of completely positive trace-preserving 
(CPTP) maps acting on the society's joint state. Table~\ref{tab_QMR1_BYH} summarizes these steps 
in a procedural form, from the input state and dephased profile through Tarjan's algorithm, 
the give the minority a shot (GMS) modification, and the final enforce-unanimity (EU) step.

\begin{table}[H]
    \centering
\begin{tabular}{|m{0.15\textwidth}|m{0.45\textwidth}|m{0.4\textwidth}|}
    \hline
    \multicolumn{3}{|c|}{QMR constitution steps} \\
    \hline
    Step & Definition& Comments\\ 
    \hline
    Input quantum state 
    &$\sigma_\text{soc}\in\mathcal{D}(\mathcal{H}_A^{\otimes n})$
    &
    $\sigma_\text{soc}$ is the society's joint quantum state\\
    \hline 
     Quantum profile (QP)&$\mathcal{P}_\text{soc}=\text{QP}(\sigma_\text{soc})=\{\rho_1,\ldots,\rho_n\}\in\mathcal{D}(\mathcal{H}_A^{\otimes n})$
    \newline 
    where $\rho_i=\text{Tr}_{V\setminus i}\,\,\sigma_\text{soc}$
    &
    $\mathcal{P}_\text{soc}$ is a $n$-tuple of the reduced, individual voters' density operators, $\rho_i$'s\\
    \hline
    Dephased profile (DP) &$\mathcal{P}_\text{soc}^d=\text{DP}(\mathcal{P}_\text{soc})=\{\rho_1^d,\ldots,\rho_n^d\}$
    \newline 
    whith $\rho_i^d= \sum_{L_i} \ket{L_i}\bra{L_i}\rho_i\ket{L_i}\bra{L_i}=\sum_L p_i^L\chi_i^L$,
    \newline
    where $p_i^L=\bra{L_i}\rho_i\ket{L_i}$, and 
    $\chi_i^L=\ket{L_i}\bra{L_i}$
    &$\mathcal{P}_\text{soc}^d$ is a $n$-tuple of the reduced and dephased,  density operators, $\rho_i^d$'s\\
    \hline
    Society's tensor product state of the dephased profile (PS)&
    $\sigma_\text{soc}^\text{DP}=\text{PS}(\mathcal{P}_\text{soc}^d)=\bigotimes_{i=1}^n\rho_i^d=
    \newline
    \sum_{L_1,\ldots,L_n}
        (p_1^{L_1}\hdots p_N^{L_n})
        (\chi_1^{L_1}\otimes\cdots\otimes\chi_N^{L_n})=
        \newline
        \sum_{\vec{L}} p(\vec{L})\vec{\chi}(\vec{L})$
        \newline 
        where 
        $\vec{L}=(L_1,...,L_N)\in\mathcal{L}(A)^n$,
        \newline
        $p(\vec{L})=p_1^{L_1}\hdots p_n^{L_n}$,
        and  
        \newline
        $\vec{\chi}(\vec{L})=\chi_1^{L_1}\otimes\cdots\otimes\chi_n^{L_n}\in\mathcal{H}_A^{\otimes n}$
        &$\sigma_\text{soc}^\text{DP}\in\mathcal{D}(\mathcal{H}_A^{\otimes n})$ is a tensor product state constructed from the dephased individual states, $\rho_i^d$'s\\
    \hline
    Digraph, SCCs' list (Tarjan's algorithm), linear extensions (DS)&
    $\chi^{(1)}(\vec{L})=
    \text{DS}\left(\vec{\chi}(\vec{L})\right)=
    \frac{1}{|\mathcal{L}_\text{SCC}|}\sum_{L\in \mathcal{L}_\text{SCC}}\ket{L}\bra{L}$
    \newline 
    where $\mathcal{L}_\text{SCC}$ represents all linear extensions.
    \newline
    \newline
    $\rho_\text{soc}^{(1)}=
    \text{DS}\left(\sigma_\text{soc}^\text{DP}\right)=
    \sum_{\vec{L}}p(\vec{L})\chi^{(1)}(\vec{L})$
    &For each $\vec{\chi}(\vec{L})\in\mathcal{D}(\mathcal{H}_A^{\otimes n})$ a directed graph is constructed. List of strongly connected components (SCC) is obtained using Tarjan's algorithm.
    The output state, $\chi^{(1)}(\vec{L})\in\mathcal{D}(\mathcal{H}_A)$, is the linear extension of the orders in the SCCs' list
    \\
    \hline
    Give the minority a shot (GMS)&
    $\chi^{(2)}(\vec{L})=
    \text{GMS}_\delta\left(\chi^{(1)}(\vec{L})\right)=\newline
    (1-\delta|k(\vec{L})|)\chi^{(1)}(\vec{L})+\delta\sum_{(a,b)\in k(\vec{L})}\frac{\Pi^{a\succ b}}{\text{Tr}(\Pi^{a\succ b})}$
    \newline
    where 
    \newline
    $k(\vec{L})=
    \newline
    \left\{(a,b):
    \text{Tr}\Big(\Pi^{a\succ b}\sigma_\text{soc}^\text{DP}\Big)>0
    \wedge
    \text{Tr}\Big(\Pi^{a\succ b}\chi^{(1)}\Big)=0
    \right\}$ is the set of all the preferences existing in $\sigma_\text{soc}^\text{DP}$ but are missing in $\chi^{(1)}(\vec{L})$.
    \newline
    $\rho_\text{soc}^{(2)}=\text{GMS}_\delta\left(\rho_\text{soc}^{(1)}\right)=
    \sum_{\vec{L}}p(\vec{L})\chi^{(2)}(\vec{L})$
    &In the GMS's output state, $\chi^{(2)}(\vec{L})$,  $\chi^{(1)}(\vec{L})$ is decreased by $\delta|k(\vec{L})|$. The states that exist in $\sigma_\text{soc}^\text{DP}$ but don't appear in $\chi^{(1)}(\vec{L})$ are added with a coefficient $\delta$, where $0 \le \delta \le 1/|k|$. This step gives a representation to minorities absent in $\chi^{(1)}(\vec{L})$\\ 
    \hline
    Enforce unanimity (EU)&
    $\chi_\text{soc}(\vec{L})=
    \text{EU}\left(\chi^{(2)}(\vec{L})\right)=
    \frac{\Pi^{U(\vec{L})}\chi^{(2)}(\vec{L})\Pi^{U(\vec{L})}}{\text{Tr}\left(\Pi^{U(\vec{L})}\chi_{(2)}(\vec{L})\right)}$
    \newline
    where $\Pi^{\text{U}(\vec{L})}=\prod_{(a,b)\in U(\vec{L})}\Pi^{a\succ b}$,
    \newline
    and $U(\vec{L})=\{(a,b):\forall i, \text{Tr}(\Pi^{\succ b}\rho^{(2)}_i(\vec{L}))=1\}$
    \newline
    $\rho_\text{soc}=
    \text{EU}_\delta\left(\rho_\text{soc}^{(2)}\right)=
    \sum_{\vec{L}}p(\vec{L})\chi_\text{soc}(\vec{L})$
    &The EU step restores the unanimity property spoiled by the GMS step\\
    \hline
    Measurement&$M_L:\rho_\text{soc}\to \ket{L}\bra{L}$&\\
    \hline
\end{tabular}
  \caption{Constitutional steps of the QMR protocol.}
    \label{tab_QMR1_BYH}
\end{table}

\subsubsection{QMR2}
The QMR2 constitution alters the aggregation step to emphasize the most frequently occurring 
individual rankings in each classical profile, and serves as a model for entanglement-enabled 
strategic behavior. Table~\ref{tab_QMR2} summarizes the QMR2 mapping from the joint 
societal state to a classical distribution over rankings.

\begin{table}[H]
    \centering
\begin{tabular}{|m{0.25\textwidth}|m{0.4\textwidth}|m{0.3\textwidth}|}
    \hline
    \multicolumn{3}{|c|}{QMR2 constitution steps} \\
    \hline
    Step & Definition& Comments\\ 
    \hline
    Input quantum state 
    &$\sigma_\text{soc}\in\mathcal{D}(\mathcal{H}_A^{\otimes n})$
    &
    $\sigma_\text{soc}$ is the society's joint quantum state\\
    \hline 
    Dephased profile (DP) &$\mathcal{P}_\text{soc}^d=\text{DP}(\mathcal{P}_\text{soc})=\{\rho_1^d,\ldots,\rho_n^d\}$
    \newline 
    with $\rho_i^d= \sum_{L_i} \ket{L_i}\bra{L_i}\rho_i\ket{L_i}\bra{L_i}=\sum_L p_i^L\chi_i^L$,
    \newline
    where $p_i^L=\bra{L_i}\rho_i\ket{L_i}$, and 
    $\chi_i^L=\ket{L_i}\bra{L_i}$
    &$\mathcal{P}_\text{soc}^d$ is a $n$-tuple of the reduced and dephased,  density operators, $\rho_i^d$'s\\
    \hline
    The number of votes in $\vec{L}$ with a specific ranking $L$ &$n_{\vec{L}}(L)=|\{i\in V:\text{Tr}(\Pi^L\rho_i^d)>0 
    \}|$ &\\
    \hline
    A preference in $\vec{L}$ with maximum representation&$M_{\vec{L}}=\underset{L\in\vec{L}}{\arg\,\max} \,\,n_{\vec{L}}(L)$&
    There might be ties, $M_{\vec{L}}$ might include more than a single $L$\\
    \hline
    Uniformly random choice in the case of ties&
    $q_{\vec{L}}(L)=
    \begin{cases} 
\frac{1}{|M_{\vec{L}}|} & L\in M_{\vec{L}}, \\
0 & \text{otherwise}.
\end{cases}
$
    &\\
    \hline
    Distribution over social (classical) preferences&
    $\pi(L)=\sum_{\vec{L}} p(\vec{L}) q_{\vec{L}}(L)$
    &\\
    \hline
    QMR2 output state&
    $\mathcal{E}_\text{QMR2}(\sigma_\text{soc})=\rho_\text{soc}^\text{QMR2}=
    \sum_{L}\pi(L)\chi^L$
    &\\
    \hline
\end{tabular}
  \caption{Constitutional steps of the QMR2 protocol.}
    \label{tab_QMR2}
\end{table}

\bibliographystyle{unsrt}
\bibliography{Q_Voting_paper.bib}

@book{vonNeumann1944,
  author    = {John von Neumann and Oskar Morgenstern},
  title     = {Theory of Games and Economic Behavior},
  publisher = {Princeton University Press},
  year      = {1944}
}

@article{Nash1951,
  author  = {John Nash},
  title   = {Non-Cooperative Games},
  journal = {Annals of Mathematics},
  volume  = {54},
  pages   = {286--295},
  year    = {1951}
}

@book{Arrow1951,
  author    = {Kenneth J. Arrow},
  title     = {Social Choice and Individual Values},
  publisher = {Yale University Press},
  year      = {1951}
}

@article{Gibbard1973,
  author  = {Allan Gibbard},
  title   = {Manipulation of voting schemes: A general result},
  journal = {Econometrica},
  volume  = {41},
  pages   = {587--601},
  year    = {1973}
}

@article{Satterthwaite1975,
  author  = {Markus Satterthwaite},
  title   = {Strategy-proofness and Arrow's Conditions: Existence and Correspondence Theorems for Voting Procedures and Social Welfare Functions},
  journal = {Journal of Economic Theory},
  volume  = {10},
  pages   = {187--217},
  year    = {1975}
}

@inproceedings{Cleve2004,
  author    = {Richard Cleve and Peter H{\o}yer and Ben Toner and John Watrous},
  title     = {Consequences and limits of nonlocal strategies},
  booktitle = {Proceedings of the 19th IEEE Annual Conference on Computational Complexity},
  pages     = {236--249},
  year      = {2004}
}

@article{Brassard2005,
  author  = {Gilles Brassard and Anne Broadbent and Alain Tapp},
  title   = {Quantum pseudo-telepathy},
  journal = {Foundations of Physics},
  volume  = {35},
  pages   = {1877--1907},
  year    = {2005}
}

@article{Meyer1999,
  author  = {David A. Meyer},
  title   = {Quantum strategies},
  journal = {Physical Review Letters},
  volume  = {82},
  pages   = {1052},
  year    = {1999}
}

@article{Eisert1999,
  author  = {Jens Eisert and Martin Wilkens and Maciej Lewenstein},
  title   = {Quantum games and quantum strategies},
  journal = {Physical Review Letters},
  volume  = {83},
  pages   = {3077},
  year    = {1999}
}

@article{CHSH1969,
  author  = {John F. Clauser and Michael A. Horne and Abner Shimony and Richard A. Holt},
  title   = {Proposed experiment to test local hidden-variable theories},
  journal = {Physical Review Letters},
  volume  = {23},
  pages   = {880},
  year    = {1969}
}

@article{BaoHalpern2017,
  author  = {Ning Bao and Nicole Yunger Halpern},
  title   = {Quantum voting and violation of Arrow's impossibility theorem},
  journal = {Physical Review A},
  volume  = {95},
  pages   = {062306},
  year    = {2017}
}

@article{Cabello2001,
  author  = {Ad{\'a}n Cabello},
  title   = {Bell's theorem without inequalities and without probabilities for two observers},
  journal = {Physical Review Letters},
  volume  = {86},
  pages   = {1911},
  year    = {2001}
}

@article{Aravind2004,
  author  = {P. K. Aravind},
  title   = {Quantum mysteries revisited again},
  journal = {American Journal of Physics},
  volume  = {72},
  pages   = {1303},
  year    = {2004}
}

@article{Xu2022,
  author  = {Jian-Ming Xu and others},
  title   = {Experimental demonstration of quantum pseudotelepathy},
  journal = {Physical Review Letters},
  volume  = {129},
  pages   = {050402},
  year    = {2022}
}

@article{Christensen2015,
  author  = {Bradley G. Christensen and Yeong-Cherng Liang and Nicolas Brunner and Nicolas Gisin and Paul G. Kwiat},
  title   = {Exploring the limits of quantum non-locality with entangled photons},
  journal = {Physical Review X},
  volume  = {5},
  pages   = {041052},
  year    = {2015}
}

@article{Tarjan1972,
  author  = {Robert E. Tarjan},
  title   = {Depth-first search and linear graph algorithms},
  journal = {SIAM Journal on Computing},
  volume  = {1},
  pages   = {146--160},
  year    = {1972}
}

@article{Sun2021,
  author  = {Xin Sun and Feiran He and Mateusz Sopek and Ming Guo},
  title   = {Schr{\"o}dinger's Ballot: Quantum Information and the Violation of Arrow's Impossibility Theorem},
  journal = {Entropy},
  volume  = {23},
  pages   = {1083},
  year    = {2021}
}

@article{HoroshkoKilin2009,
  author  = {Dmitri Horoshko and Sergei Kilin},
  title   = {Quantum anonymous voting with anonymity check},
  journal = {Physics Letters A},
  volume  = {375},
  number  = {8},
  pages   = {1172--1175},
  year    = {2011}
}

@article{Xu2022QuantumVotingNoMemory,
  author  = {Lidong Xu and Mingqiang Wang},
  title   = {Quantum voting protocol without quantum memory},
  journal = {Frontiers in Physics},
  volume  = {10},
  pages   = {1023992},
  year    = {2022}
}

@article{Xiong2022SingleParticleVoting,
  author  = {Zihao Xiong and Aihan Yin},
  title   = {Single particle electronic voting scheme based on quantum ring signature},
  journal = {Modern Physics Letters A},
  volume  = {37},
  number  = {26},
  pages   = {2250174},
  year    = {2022}
}

@article{QuantumLogicalVote2022,
  author  = {Xin Sun and Feifei He and Daowen Qiu and Piotr Kulicki and Mirek Sopek and Meiyun Guo},
  title   = {Distributed Quantum Vote Based on Quantum Logical Operators, a New Battlefield of the Second Quantum Revolution},
  journal = {arXiv preprint arXiv:2202.00147},
  year    = {2022}
}

@article{LiuHanXiaYu2023AcceleratingVoting,
  author  = {Ao Liu and Qishen Han and Lirong Xia and Nengkun Yu},
  title   = {Accelerating Voting by Quantum Computation},
  journal = {Proceedings of Machine Learning Research},
  volume  = {216},
  pages   = {1274--1283},
  year    = {2023}
}

@article{Qiskit,
  author       = {Schuld, Maria and Cross, Andrew W. and McKay, David C. and
                  Alexander, Thomas and Abraham, H{\'e}ctor and Akhalwaya, Ismail Yunus and
                  Aleksandrowicz, Gadi and others},
  title        = {Qiskit: An Open-source Framework for Quantum Computing},
  journal      = {Journal of Open Source Software},
  year         = {2023},
  volume       = {8},
  number       = {86},
  pages        = {5131},
  doi          = {10.21105/joss.05131},
  note         = {Qiskit version used in this work includes noisy simulators and FakeBackends.}
}

\end{document}